\documentclass[prc,aps,floatfix,showpacs,twocolumn]{revtex4}
\usepackage{dcolumn}
\usepackage{bm}
\usepackage{color}
\usepackage{graphicx}
\usepackage[active]{srcltx}
\usepackage{pstricks}
\def\m{{\phantom{-}}}
\begin{document}

\title{The parity-violating asymmetry in the $^3$He($\vec{n},p$)$^3$H reaction}
\author{M.\ Viviani$^{\,{\rm a}}$, R.\ Schiavilla$^{\,{\rm b,c}}$,
L.\ Girlanda$^{\,{\rm d,a}}$, A.\ Kievsky$^{\,{\rm a}}$, and L.E.\ Marcucci$^{\,{\rm  d,a}}$
}
\affiliation{
$^{\,{\rm a}}$\mbox{INFN-Pisa, 56127 Pisa, Italy}\\
$^{\rm b}$\mbox{Department of Physics, Old Dominion University, Norfolk, VA 23529, USA} \\
$^{\rm c}$\mbox{Jefferson Lab, Newport News, VA 23606}\\
$^{\rm d}$\mbox{Department of Physics, University of Pisa, 56127 Pisa, Italy}\\
}

\date{\today}

\begin{abstract}
The longitudinal asymmetry induced by parity-violating (PV) components in the nucleon-nucleon
potential is studied in the charge-exchange reaction $^3$He($\vec{n},p$)$^3$H at vanishing
incident neutron energies.  An expression for the PV observable is derived in terms of $T$-matrix
elements for transitions from the $^{2S+1}L_J=\, ^1$S$_0$ and $^3$S$_1$ states in the
incoming $n$-$^3$He channel to states with $J=0$ and 1 in the outgoing $p$-$^3$H channel.
The $T$-matrix elements involving PV transitions are obtained in first-order perturbation theory
in the hadronic weak-interaction potential, while those connecting states of the same parity are
derived from solutions of the strong-interaction Hamiltonian with the hyperspherical-harmonics
method.  The coupled-channel nature of the scattering problem is fully accounted for.  Results are
obtained corresponding to realistic or chiral two- and three-nucleon strong-interaction potentials
in combination with either the DDH or pionless EFT model for the weak-interaction potential.
The asymmetries, predicted with PV pion and vector-meson coupling constants corresponding
(essentially) to the DDH ``best values'' set, range from --9.44 to --2.48 in units of $10^{-8}$,
depending on the input strong-interaction Hamiltonian.  This large model dependence is a
consequence of cancellations between long-range (pion) and short-range (vector-meson)
contributions, and is of course sensitive to the assumed values for the PV coupling constants.
\end{abstract}


\maketitle

\section{Introduction, results, and conclusions}
\label{sec:intro}

A number of experiments aimed at studying parity violation in low-energy processes
involving few nucleon systems are being completed or are in an advanced stage of
planning at cold neutron facilities, such as the Los Alamos Neutron Science Center,
the NIST Center for Neutron Research, and the Spallation Neutron Source at Oak Ridge.
The primary objective of this program is to determine the fundamental parameters
of hadronic weak interactions, in particular the strength of the long-range part
of the parity-violating (PV) two-nucleon ($NN$) potential, mediated by one-pion
exchange (OPE).  While such a component is theoretically expected on the basis of
the weak interactions between quarks and the spontaneously-broken chiral symmetry
of QCD, experimental evidence for its presence has proven to be elusive, and indeed
current constraints are inconclusive, for a review see Ref.~\cite{Haxton08}.

In contrast, in the strong-interaction sector OPE dominates the $NN$ potential
at internucleon separations larger than 1.5 fm, and the spatial-spin-isospin
correlations it induces leave their imprint on many nuclear properties.  These
include, for example, i) the observed ordering of levels in light nuclei and, in
particular, the observed absence of stable systems with mass number $A$~=~8~\cite{Wiringa02},
ii) the single-particle energy spacings and shell structure of medium- and
heavy-weight nuclei~\cite{Otsuka05}
and, in particular, the observed change in the energy gap between the $h_{11/2}$ and
$g_{7/2}$ orbits in tin isotopes~\cite{Schiffer04}, and iii) the relative magnitude
of the momentum distributions of $pp$ versus $np$ pairs in nuclei~\cite{Schiavilla07},
which leads to the strong suppression of $(e,e^\prime pp)$ relative to $(e,e^\prime np)$
knock-out cross sections from $^{12}$C, recently measured at Jefferson Lab~\cite{Subedi08}.

The determination of the parameters that characterize parity violation in nuclei
requires evaluating matrix elements of hadronic weak-interaction operators between
eigenstates of the strong-interaction Hamiltonian.  Thus, experiments in this field
are especially reliant on theory for their analysis and interpretation.  For this
reason, over the last several years, we have embarked on a program aimed at
developing a systematic framework for studying PV observables in few-nucleon
systems, for which accurate---essentially exact---calculations are possible.  Two
earlier papers~\cite{Carlson02,Schiavilla04} dealt with the two-nucleon system,
and provided a rather complete analysis of the longitudinal asymmetry in $\vec{p}$-$p$
scattering~\cite{Carlson02} up to 300 MeV lab energies, and of a variety of PV
observables in the $np$ system~\cite{Schiavilla04}, including, among others, the
neutron spin rotation in $\vec{n}$-$p$ scattering and the photon angular asymmetry
in the $\vec{n}$-$p$ radiative capture at thermal neutron energies.  In the next
phase, we have studied the spin rotation in $\vec{n}$-$d$~\cite{Schiavilla08}
and $\vec{n}$-$\alpha$~\cite{Arriaga10} scattering at cold neutron energies.

Measurements are available for the following PV observables: the longitudinal analyzing
power in $\vec{p}$-$p$~\cite{Balzer80}--\cite{Berdoz03} and $\vec{p}$-$\alpha$~\cite{Lang85}
scattering, the photon asymmetry and photon circular polarization in, respectively, the
$^1$H($\vec{n},\gamma$)$^2$H~\cite{Cavaignac77}--\cite{Gericke09}
and $^1$H($n,\vec{\gamma}$)$^2$H~\cite{Knyazkov84} radiative captures, and the neutron
spin rotation in $\vec{n}$-$\alpha$ scattering~\cite{Snow09,Bass09}.  There is also a set
of experiments which are currently being planned, including measurements of the neutron spin
rotation in $\vec{n}$-$p$~\cite{Snow09} and $\vec{n}$-$d$~\cite{Markoff07} scattering,
and of the longitudinal asymmetry in the charge-exchange reaction $^3$He($\vec{n},p$)$^3$H
at cold neutron energies~\cite{Bowman07}, the subject of the present paper.

At vanishing neutron energies, the only channels entering the incoming $n$-$^3$He scattering state
have quantum numbers $^{2S+1}L_J =  \,^1$S$_0$ and $^3$S$_1$.  In the outgoing $p$-$^3$H
scattering state, the relevant channels are: $^{2S+1}L_J = \, ^1$S$_0$, $^3$S$_1$, $^3$D$_1$
with positive parity, and $^3$P$_0$, $^1$P$_1$, $^3$P$_1$ with negative parity.  We show (in Sec.~\ref{sec:form})
that the PV observable in this process, {\it i.e.}~the longitudinal analyzing power $A_z$, reads
\begin{equation}
A_z= a_z \, {\rm cos}\, \theta \ ,
\label{eq:aazz}
\end{equation}
where $\theta$ is the angle between the proton momentum and the neutron beam direction, and
the coefficient $a_z$ can be expressed in terms of products of $T$-matrix elements involving
(three) parity-conserving (PC) and (three) PV transitions as
\begin{widetext}
\begin{equation}
 a_z = -{4\over \Sigma}{\rm Re}
\Big( \sqrt{3}\, \overline T^{21,1}_{01,10}\,
               \overline T^{21,0\, *}_{00,00}-
\overline T^{21, 0 }_{00,11}\,
                \overline T^{21,1\, *}_{01,01}
 +\sqrt{2}\,
\overline T^{21,0 }_{00,11}\,
                \overline T^{21,1\, * }_{01,21}
+\sqrt{6}\, \overline T^{21,1}_{01,11}\,
               \overline T^{21,1\, *}_{01,01}
+\sqrt{3}\, \overline T^{21,1 }_{01,11}\,
                \overline T^{21,1\, *}_{01,21}  \Big)\  ,
\label{eq:aaz1} 
\end{equation}
\end{widetext}
and
\begin{equation}
\Sigma= \left|\overline{T}^{21, 0}_{00,00}\right |^2 + 
            3\,\left |\overline{T}^{21, 1}_{01,01}\right |^2 + 3\, \left |\overline{T}^{21,1}_{01,21}\right |^2 \ .
\label{eq:ssss}
\end{equation}
In $\overline{T}^{21,J}_{LS,L'S'}$ the label $J$ specifies the total angular momentum, the superscripts
$21$ denote the charge-exchange transition $n$-$^3$He to $p$-$^3$H (as opposed, for
example, to the elastic transition, which would be denoted by the superscripts $22$), the subscripts
$LS$ ($L^\prime S^\prime$) are the relative orbital angular momentum and channel spin of the
$n$-$^3$He ($p$-$^3$H) clusters, and lastly the overline is to note the inclusion of a convenient
phase factor---see Eq.~(\ref{eq:tbar}) below.  The PC (PV) $T$-matrix elements have $L+L^\prime$
even (odd), and the sum $\Sigma$ in Eq.~(\ref{eq:ssss}) is proportional to the $^3$He($n,p$)$^3$H cross section.
We observe that $a_z$ vanishes if only the channels $^1$S$_0$ and $^3$P$_0$ (with $J=0$) are retained.

The $T$-matrix elements are related to the (real) $R$-matrix elements
(Sec.~\ref{sec:tma} and Appendix~\ref{app:a1}),
and the latter for PC transitions are calculated via the Kohn variational principle with the
hyperspherical-harmonics (HH) method~\cite{Kievsky08,Viviani10}
(Sec.~\ref{sec:whh}).  We use strong-interaction 
Hamiltonian models, consisting of the Argonne $v_{18}$ (AV18)~\cite{Wiringa95} or chiral
(N3LO)~\cite{Entem03} two-nucleon potential in combination with the Urbana IX (UIX)~\cite{Pudliner97}
or chiral (N2LO)~\cite{Navratil07} three-nucleon potential.  The HH calculation is a challenging one,
for two reasons.  The first is the coupled-channel nature of the scattering problem: even at
vanishing energies for the incident neutron, the elastic $n$-$^3$He and charge-exchange $p$-$^3$H
channels are both open.  The second is the presence of a $J^\pi=0^+$ resonant state (of zero
total isospin) between the $p$-$^3$H and $n$-$^3$He thresholds, which slows down the
convergence of the expansion, and requires a large number of
HH basis functions in order to achieve numerically stable results. Further discussion of
this aspect of the calculations is in Sec.~\ref{sec:whh}, where we also present current predictions
for the $n$-$^3$He scattering lengths corresponding to the Hamiltonian models mentioned earlier.
They are in good agreement with the measured values.

The $R$-matrix elements involving PV transitions are computed in first-order perturbation
theory with Quantum Monte Carlo techniques (Sec.~\ref{sec:calc}).  We adopt as PV potential
the meson-exchange (DDH) model of Desplanques {\it et al.}~\cite{Desplanques80} as well as the
pionless effective-field-theory (EFT) model recently derived in Refs.~\cite{Zhu05,Girlanda08}
(Sec.~\ref{sec:pv}), and present results for the various components of the DDH and EFT potentials
in combination with the AV18, AV18/UIX, N3LO, and N3LO/N2LO Hamiltonians
in Sec.~\ref{sec:res}.  Additional results for the $R$- and $T$-matrix elements, and
combinations thereof entering the PV observable, are listed (for the AV18/UIX) in Appendix~\ref{app:a2}
for completeness.  For the DDH model only, we also present predictions for $a_z$
corresponding essentially---but see Sec.~\ref{sec:pv} for further details---to the 
``best values'' of the $\pi$-, $\rho$-, and $\omega$-meson weak-interaction coupling
constants~\cite{Desplanques80}.  These predictions range from --9.44 to --2.48 in units of $10^{-8}$
depending on whether the N3LO/N2LO or AV18/UIX Hamiltonian is considered, and thus exhibit
a significant model dependence due to cancellations (or lack thereof) between
the pion and vector-meson contributions.

It is useful to express the asymmetry as
\begin{equation}
a_z = h_\pi^1 \, C_\pi^1 +h_\rho^0 \, C_\rho^0+h_\rho^1 \, C_\rho^1+
h_\rho^2 \, C_\rho^2 +h_\omega^0\, C_\omega^0
+h_\omega^1\, C_\omega^1 \ ,\label{eq:cddh}
\end{equation}
where the $h^i_\alpha$'s, $\alpha=\pi$, $\rho$, $\omega$ and $i=0$, 1, 2,
denote the PV coupling constants in the DDH model along with the isospin
content of the corresponding interaction.  The coefficients $C_\alpha^i$ are
listed in Table~\ref{tb:tcs}, and depend on the input Hamiltonian used
to generate the continuum wave functions, as well as on the assumed
values for the PC pion- and vector-meson coupling constants
and associated cutoffs (see Table~\ref{tb:tabpv}).
\begin{widetext}
\begin{center}
\begin{table}[bth]
\begin{tabular}{c|c||c|c|c||c|c}
\hline
\hline
  &  $C_\pi^1$  &  $C_\rho^0$  &  $C_\rho^1$  & $C_\rho^2$ & $C_\omega^0$  &$C_\omega^1$  \\
\tableline
 AV18  & --0.1892(86)   & --0.0364(40)  &+0.0193(9)    & --0.0006(1)   &--0.0334(29)   &  +0.0413(10)    \\
 AV18/UIX &--0.1853(150) &--0.0380(70)    & +0.0230(18)   &  --0.0011(1)  &--0.0231(56)    &   +0.0500(20) \\
N3LO  &  --0.1989(87)  &   --0.0120(49) & +0.0242 (9)    &  +0.0002(1)  & +0.0080(30)  &   +0.0587(11)   \\
N3LO/N2LO  & --0.1110(75)   &  +0.0379(56)  & +0.0194 (10)  &   --0.0007(1) & +0.0457(36)   &   +0.0408(14)   \\
\hline
\hline
\end{tabular}
\caption{The coefficients $C_\alpha^i$ entering the PV observable $a_z$, corresponding to the
AV18, AV18/UIX, N3LO, and N3LO/N2LO strong-interaction Hamiltonians.  The statistical
errors due to the Monte Carlo integrations are indicated in parentheses, and correspond
to a sample consisting of $\sim 130$k configurations.}
\label{tb:tcs}
\end{table}
\end{center}
\end{widetext}
The coefficients $C_\alpha^i$ follow from the linear combination given in Eq.~(\ref{eq:aaz1}).
Isotensor $\rho$-exchange ($C_\rho^2$) is negligible.  The
isoscalar and isovector vector-meson exchanges give contributions of the same magnitude,
both of which are smaller than OPE.  However, the OPE contribution seems to
be significantly suppressed.  For example, in the case of the neutron spin rotation in
$\vec{n}$-$d$ scattering this contribution is calculated to be at least a factor of
$\sim 30$ larger than that of any of the $\rho$ and $\omega$ exchanges, which is not the case
for the process under consideration.  This may be due to the predominant
isoscalar character of the $^1$S$_0$ and $^3$P$_0$ channels---see discussion
in Appendix~\ref{app:a2}.  The N3LO/N2LO results should be considered as preliminary,
since the HH solution for the $0^+$ wave function has not yet fully converged
(at least as far as the singlet scattering length is concerned, see Sec.~\ref{sec:whh}).
This fact may explain why the inclusion of a three-nucleon
potential like N2LO~\cite{Navratil07} should reduce $C_\pi^{1}$ by almost a
factor of two relative to the other models.
This point will be discussed in Secs.~\ref{sec:whh} and~\ref{sec:calc}.
Finally we note that the ``best values'' for the PV couplings constants of the
pion and $\rho$-meson are (in units of $10^{-7}$) respectively +4.56 and
--16.4, and this leads to the large cancellation (and consequent model
dependence) in the values predicted for $a_z$ and referred to earlier.

We conclude by observing that the EFT analysis presented in this work
could be improved by employing chiral potentials in both the strong-
and weak-interaction sectors.  At order $Q/\Lambda_\chi$, where $Q$ is the
low energy/momentum scale that characterizes the particular process of
interest, and $\Lambda_\chi \simeq 1$ GeV is the chiral-symmetry-breaking scale,
the PV potential contains 7 low-energy constants (LECs), 5 of which are
associated with four-nucleon contact terms, and the remaining 2 with
long-range OPE components~\cite{Zhu05}.  When electromagnetic interactions
are also introduced, another (unknown) LEC must be included---it is needed to fix the strength
of a PV two-body current operator of pion range~\cite{Zhu05}.  One can
envisage, at least in principle, a suite of experiments involving $A=2$--5
systems, which would constrain, in fact over-constrain, these eight LECs.
Some of these have been mentioned above, additional ones include,
for example, measurements of the photon asymmetries in the radiative
captures $^2$H($\vec{n},\gamma$)$^3$H and $^3$He($\vec{n},\gamma$)$^4$He.
These processes are strongly suppressed: the experimental
values for the corresponding (PC) cross sections~\cite{Jurney82,Wolfs89} are, respectively,
almost 3 and 4 orders of magnitude smaller than measured in
$^1$H($n,\gamma$)$^2$H.  One would naively expect relatively
large PV asymmetries in these cases, possibly orders of magnitude larger than in the
$A$=2 system.  Clearly, accurate theoretical estimates for them
could be useful in motivating our experimental colleagues to carry out these
extremely challenging measurements.

From a theoretical perspective, most of the methodological
and technical developments needed to carry out the calculations are
already in place.  We have recently reported results~\cite{Girlanda10} for the
$A=3$ and 4 (PC) captures, using wave functions obtained from the N3LO/N2LO
Hamiltonian and electromagnetic currents derived in chiral EFT up to one
loop~\cite{Pastore09}, which are in excellent agreement with data.  However,
there is one aspect in the computation of the proposed PV threshold captures, which
still needs to be addressed: the determination of the small admixtures induced
by the PV potential into the bound and continuum wave functions.  Even a
first-order perturbative treatment of those admixtures requires construction
of the full Green's function for the strong (PC) Hamiltonian, an impractical
task.  However, it may be possible to generate them using correlated
basis methods, similar to those employed in Ref.~\cite{Schiavilla87}.

\section{The parity-violating observable}
\label{sec:form}

The neutron energies in the reaction $^3$He($\vec{n},p$)$^3$H
of interest here are in the meV range, and at these energies only
two channels are open: the $n$-$^3$He elastic channel and the $p$-$^3$H
charge-exchange channel.  In the following,
the index $\gamma$=1 (2) is used to identify the $p$-$^3$H ($n$-$^3$He)
clusters in the final (initial) state.  In the absence of strong and weak interactions
between the two clusters,
the wave function in channel $\gamma$ is written as
\begin{eqnarray}
\Phi_{\gamma}^{m_3 \, m_1}&=&{1\over\sqrt{4}} \sum_{p=1}^4 \Psi_{\gamma}^{m_3}(ijk)
   \chi_{\gamma}^{m_1}(l)\, \phi_{{\bf q}_\gamma}({\bf y}_p) \nonumber \\
&\equiv& {1\over\sqrt{4}} 
\sum_{p=1}^4 \Phi_{\gamma,\, p}^{m_3 \, m_1}\ ,
  \label{eq:pw}
\end{eqnarray}
where $\Psi_\gamma^{m_3}$ is the (antisymmetrized) trinucleon
bound-state wave function in spin projection $m_3$, $\chi_\gamma^{m_1}$
is the nucleon spin-isospin state with spin and isospin projections $m_1$
and $p$ for $\gamma$=1 or $n$ for $\gamma$=2, respectively, and
$\phi$ is the inter-cluster wave function, {\it i.e.}~a Coulomb
wave function for $\gamma=1$ or simply a plane wave $e^{i\,{\bf q}_2 \cdot {\bf y}_p}$
for $\gamma=2$.  The separation between the center-of-mass positions
of the two clusters is denoted by ${\bf y}_p$ with ${\bf y}_p ={\bf r}_l- {\bf R}_{ijk}$, and
their relative momentum is specified by ${\bf q}_\gamma$, so that the energy
$E$ is given by
\begin{equation}
 E=-B_\gamma+{q_\gamma^2\over 2\, \mu_\gamma}\ , 
\qquad {1\over \mu_\gamma}= {1\over m_\gamma}+{1\over M_\gamma}\ .\label{eq:ecm}
\end{equation}
Here $B_\gamma$ and $M_\gamma$ are the binding energy and mass
of $^3$H ($^3$He) for $\gamma$=1 (2), and $m_\gamma$ is the proton
(neutron) mass for $\gamma$=1 (2).  Lastly, the wave functions in Eq.~(\ref{eq:pw})
are antisymmetrized by summing over the four permutations $p$ with
$(ijk,l)\equiv (123,4)$, $(124,3)$, $(134,2)$, and $(234,1)$.

It is useful to expand the wave functions in Eq.~(\ref{eq:pw}) in partial waves as
\begin{equation}
  \Phi_\gamma^{m_3\, m_1}= {1\over\sqrt{4}} \sum_{p=1}^4
\sum_{LSJ} i^L\, Z_{m_3 \, m_1}^{L0SJJ_z}\,\, \Omega^{JJ_z}_{\gamma LS,\, p}\,\,
\frac{{\cal F}^F_L(q_\gamma; y_{p})}{q_\gamma y_p}\ ,
\label{eq:pw1}
\end{equation}
where ${\cal F}^F_L(q_\gamma; y_{p})$ reduces to a regular Coulomb function $F_L(q_\gamma; y_{p})$
(multiplied by a phase factor we need not specify here) for $\gamma=1$
or a spherical Bessel function $x\,j_L(x)$ for $\gamma=2$,
with $x=q_\gamma y_p$.  The channel
functions $ \Omega^{JJ_z}_{\gamma LS,\, p}$ are defined as
\begin{equation}
\Omega^{JJ_z}_{\gamma LS,\, p}=\Big [ Y_L(\hat{\bf y}_p)\otimes 
\big[ \Psi_\gamma (ijk) \otimes \chi_\gamma(l) \big]_S \Big]_{JJ_z}\ ,
\label{eq:omega1}
\end{equation}
while the Clebsch-Gordan coefficients associated with the re-coupling of
the angular momenta (and other factors) are lumped into
\begin{eqnarray}
 Z_{m_3\, m_1}^{LMSJJ_z}& =& \sqrt{4\pi}\, \sqrt{2L+1}\, 
\langle 1/2, m_3\ ;1/2, m_1 | S , S_z\rangle \nonumber \\
&&\qquad\qquad\qquad \times \langle L, M; S, S_z|J, J_z\rangle \ .
\end{eqnarray}
The momentum ${\bf q}_\gamma$ has been taken to define
the spin-quantization axis, {\it i.e.}~the $z$-axis.

In the presence of inter-cluster interactions, the $n$-$^3$He wave function in the asymptotic region
reads
\begin{widetext}
\begin{eqnarray}
 \Psi_{\gamma=2}^{m_3\, m_1}&\simeq&   {1\over\sqrt{4}}\sum_{p=1}^4 \sum_{LSJ}
 i^L\, Z_{m_3 \,  m_1}^{L0SJJ_z}\,
        \Biggl[\Omega^{JJ_z}_{2 LS,\, p}\, j_L(q_2 y_{p})
+\sum_{L'S'} T^{22, J}_{LS,\,L'S'} \Omega^{JJ_z}_{2 L'S',\, p}\; {e^{i(q_2
        y_{p}-L'\pi/2)} \over y_{p}} \nonumber \\
&&\qquad +\sum_{L'S'} 
      T^{21,  J}_{LS,\, L'S'} \Omega^{JJ_z}_{1 L'S',\, p} \; {e^{i[q_1 y_{p}-L'\pi/2-\eta_1\ln(2\, q_1
        y_{p})+\sigma_{L'}]} \over y_{p}} \Biggr]\ ,
\label{eq:dw}
\end{eqnarray}
\end{widetext}
and contains outgoing spherical waves in the $n$-$^3$He elastic channel ($\gamma=2$)
as well as in the $p$-$^3$H charge-exchange channel ($\gamma=1$) multiplied by
corresponding $T$-matrix elements $T^{\, \gamma \gamma', J}_{LS,L'S'}$.  Here
$\eta_1=\alpha\mu_1/q_1$, where $\alpha$ is the fine structure constant and $\mu_1$ is
the $p$-$^3$H reduced mass defined above, and $\sigma_L$ is the Coulomb phase-shift.  Thus
Coulomb distortion in the $p$-$^3$H outgoing state is fully accounted for.

The probability amplitude $M_{m_3' \,m_1', \, m_3 \, m_1}$  to observe a $p$-$^3$H
final state with spin projections $m_1'$ and $m_3'$, respectively, is obtained from
\begin{eqnarray}
\langle \Phi^{m_3'\, m_1'}_{\gamma=1,\, p=1}\mid \Psi_{\gamma=2}^{m_3\, m_1}\rangle
={1\over\sqrt{4}} M_{m_3'\, m_1', \, m_3\, m_1} \nonumber \\
\times {e^{i[q_1 y-\eta_1\ln(2\, q_1 y)]} \over y}\ ,
\label{eq:mmat}
\end{eqnarray}
where we have assumed that the $p$-$^3$H state is in partition (123,4) corresponding
to permutation $p=1$, namely the bound cluster consists of particles 123 and
the proton is particle 4.  For brevity, we have also set
${\bf y}\equiv {\bf y}_{p=1}$.  Using the orthonormality of the channel functions
$\Omega^{JJ_z}_{\gamma LS,\, p}$, we find
\begin{eqnarray}
M_{m_3'  m_1', \, m_3 \, m_1}\!\!&=&\!\!\frac{1}{\sqrt{4\pi}}\!\!
   \sum_{JLS L'S'} i^L (-i)^{L'}  \frac{e^{i\,\sigma_{L'}}}{\sqrt{2L'+1}}\,
Z_{m_3 \,  m_1}^{L0SJJ_z}\nonumber \\
&&\times T^{21, J}_{LS,\, L'S'}\, Z_{m_3' \,  m_1'}^{L'M'S'JJ_z}\,
Y_{L'M'}(\hat{\bf y})\ ,
\label{eq:mmat1}
\end{eqnarray}
where the Clebsch-Gordan coefficients require
$J_z=S_z=m_3+m_1$, $S_z'=m_3'+m_1'$, and $M'=J_z-S_z'= m_3+m_1- (m_3'+m_1')$.

The spin-averaged cross section follows from
\begin{equation}
 \sigma_0\equiv \frac{{\rm d}\sigma}{{\rm d}\Omega}={1\over 4}
{ \mu_2 \over \mu_1} {q_1\over q_2}
 \sum_{m_3,m_1} \sum_{m_3', m_1'} 
| M_{m_3'\, m_1',\, m_3\, m_1}|^2 \ ,  \label{eq:ucs}
\end{equation}
since $(1/4) (q_1/\mu_1)\, |M_{m_3' \, m_1',\, m_3\ ,m_1}|^2 \,{\rm d} \Omega$ is the flux of
outgoing particles in the solid angle \hbox{${\rm d}\Omega\equiv {\rm d}\hat{\bf y}$},
and $(1/4)(q_2/\mu_2)$ is the incident flux, where the factors $1/4$ originate from the normalization factors
$1/\sqrt{4}$ in Eqs.~(\ref{eq:pw}) and (\ref{eq:mmat}).   These cancel out in Eq.~(\ref{eq:ucs}),
leaving an extra $1/4$ coming from the average over the initial polarizations.
The longitudinal asymmetry $A_z$ is defined as
\begin{eqnarray}
 \sigma_0 \, A_z&=&{1\over 2} { \mu_2 \over \mu_1} {q_1\over q_2}
\sum_{m_3} \sum_{m_3', m_1'} \biggl[
   | M_{m_3'\, m_1',\,m_3\, m_1=+{1\over 2}}|^2 \nonumber \\
&&\qquad\qquad - | M_{m'_3\, m_1', \,m_3\,m_1=-{1\over 2}}|^2 \biggr]
 \ .
\label{eq:az}
\end{eqnarray}

At meV energies it suffices to keep only $L=0$ in the entrance channel, so that
\begin{eqnarray}
   M_{m_3'\, m_1', \, m_3\, m_1}\!\!\!&=&\!\!\!
   \sum_{J=0,1} \sum_{L'S'} 
    \langle 1/2, m_3;1/2, m_1 | J , J_z\rangle \nonumber \\
\!\!&\times &\!\!\frac{\overline{T}^{21,J}_{0J,L'S'}}{\sqrt{2L'+1}}
Z_{m_3' \,  m_1'}^{L'M'S'JJ_z} Y_{L'M'}(\hat{\bf y})        \ ,
\label{eq:m1}
\end{eqnarray}
where we have defined
\begin{equation} 
   \overline T^{21,J}_{0J,L'S'}= (-i)^{L'}\, e^{i\, \sigma_{L'}}\,
   T^{21, J}_{0J,L'S'}\ . \label{eq:tbar}
\end{equation}
After inserting the expression for $Z_{m_3' \,  m_1'}^{L'M'S'JJ_z}$ and carrying out
the sums over $m_1, m_3$ and $m_1',m_3'$, we find the unpolarized cross section
to be given by
\begin{eqnarray}
  \sigma_0\!\!&=&\!\! {1\over 4}  { \mu_2 \over \mu_1} {q_1\over q_2} 
\sum_{J=0,1}\sum_{L'S'} (2J+1) \left |\overline T^{21,J}_{0J,L'S'}\right|^2 \nonumber \\
\!\!&=& \!\!{1\over 4}  { \mu_2 \over \mu_1} {q_1\over q_2} 
\Bigg[\left|T^{21, 0}_{00,00}\right |^2\!\! +
            3\! \left |T^{21, 1}_{01,01}\right |^2 \!\!+ 3\! \left |T^{21,1}_{01,21}\right |^2\Bigg] \ ,
 \label{eq:ucs1}
\end{eqnarray}
where in the second line we have ignored $T$-matrix elements involving
transitions to odd parity final states (and hence parity violating), since
these are induced by hadronic weak interactions and consequently are
much smaller than the parity-conserving $T$-matrices associated with strong
interactions.  We observe that the matrix elements $T^{21,J}$ (and $\overline T^{21,J}$)
are finite in the limit $q_2=0$, and therefore $\sigma_0$ is divergent as  $q_2$
goes to zero, as expected for a neutron capture reaction.

\begin{table}[bth]
\caption{The coefficients $C^{J_1 J_2}_{L_1 L_2 S}(|M|)$ for the relevant channels.}
  \begin{tabular}{l| llll|l} 
\hline
\hline
   $J_1,J_2$ &   $L_1$ & $L_2$ & $S$ & $|M|$ & $ C^{J_1 J_2}_{L_1 L_2 S}(|M|)$ \\
\hline
 $0,1$ &  $0$ &   $1$  & $0$ & $0$ & $-\sqrt{3}$ \\
      &  $1$ &   $0$  & $1$ & $0$ & $+1$ \\
     &  $1$ &   $2$  & $1$ & $0$ & $-\sqrt{2}$ \\
     &  $1$ &   $2$  & $1$ & $1$ & $-\sqrt{1/2}$ \\
\hline
 $1,0$ &  $1$ &   $0$  & $0$ & $0$ & $-\sqrt{3}$ \\
    &  $0$ &   $1$  & $1$ & $0$ & $+1$ \\
    &  $2$ &   $1$  & $1$ & $0$ & $-\sqrt{2}$ \\
    &  $2$ &   $1$  & $1$ & $1$ & $-\sqrt{1/2}$ \\
\hline
 $1,1$ &  $0$ &   $1$  & $1$ & $0$ & $-\sqrt{6}$ \\
    &  $2$ &   $1$  & $1$ & $0$ & $-\sqrt{3}$ \\
    &  $2$ &   $1$  & $1$ & $1$ & $-\sqrt{3/4}$ \\
    &  $1$ &   $0$  & $1$ & $0$ & $-\sqrt{6}$ \\
    &  $1$ &   $2$  & $1$ & $0$ & $-\sqrt{3}$ \\
    &  $1$ &   $2$  & $1$ & $1$ & $-\sqrt{3/4}$ \\
\hline
\hline
\end{tabular}
\label{tb:tabc}
\end{table}

The asymmetry $A_z$ can be written as
\begin{eqnarray}
 \sigma_0 \, A_z \!\!&=&\!\! \frac{1}{2}{ \mu_2 \over \mu_1} {q_1\over q_2}\!\!
   \sum_{J_1,J_2=0,1}\sum_{L_1L_2S}\!\! \epsilon_{L_1 L_2}  \overline
   T^{21,J_1}_{0J_1,L_1S} 
    \left[\overline  T^{21,J_2}_{0J_2,L_2S}\right]^* \nonumber \\
&\times& \sum_{|M|} 
         C^{J_1 J_2}_{L_1L_2S}(|M|)
        P^{|M|}_{L_1}(\theta)  P^{|M|}_{L_2}(\theta)
        \ ,\label{eq:d2}
\end{eqnarray}
where the $P_L^{|M|}(\theta)$'s are associated Legendre functions, $\theta$ is
the angle of the outgoing proton momentum relative to the direction of the incident
beam, the $C^{J_1 J_2}_{L_1L_2S}(|M|)$'s denote combinations
of Clebsch-Gordan coefficients, defined as
\begin{eqnarray}
  C^{J_1 J_2}_{L_1 L_2S}(|M|)\!\!&=&\!\!\frac{1}{2 \pi}
      \sum_{J_z}  \sum_{\mu =\pm |M|}\! \sqrt{(L_1\!-\!|M|)! (L_2\!-\!|M|)!\over
      (L_1\!+\!|M|)! (L_2\!+\!|M|)!}\nonumber \\  
&&\times Z_{m_3 \,  m_1=+1/2}^{L_1 \mu\, SJ_1 J_z}\,
Z_{m_3 \,  m_1=+1/2}^{L_2 \mu\, SJ_2 J_z} \ ,
\label{eq:ccoef}
\end{eqnarray}
and lastly the phase factor $\epsilon_{L_1 L_2}$,
\begin{equation}
\epsilon_{L_1 L_2}\equiv{1-(-)^{L_1+L_2}\over 2} \ ,
\end{equation}
ensures that either $L_1$ or $L_2$ must be odd, which in turn implies
that either $T^{21,J_1}_{0J_1,L_1S}$ or $T^{21,J_2}_{0J_2,L_2S}$ involves
a parity-violating transition, {\it i.e.}~a transition from an incoming positive parity
$n$-$^3$He state to an outgoing negative parity $p$-$^3$H state.  The non-vanishing
$C$'s  for the relevant channels are listed in Table~\ref{tb:tabc}, and evaluation of the
sums in Eq.~(\ref{eq:d2}) allows one to express the parity-violating asymmetry as
in Eqs.(\ref{eq:aazz})--(\ref{eq:ssss}).

\section{$T$-matrix elements}
\label{sec:tma}

The calculation proceeds in two steps: we first determine, via the Kohn variational principle,
the $R$-matrix elements, and then relate these to the $T$-matrix
elements.  The wave function describing a scattering state with total
angular momentum $JJ_z$ in channel $\gamma LS$ is written as
\begin{equation}
     \Psi^{JJ_z}_{\gamma,LS}=\Psi^{C,JJ_z}_{\gamma,LS}+\Psi^{F,JJ_z}_{\gamma,LS}+ \sum_{\gamma' L'S'}
      R^{\gamma \gamma', J}_{LS,L'S'} \Psi^{G,JJ_z}_{\gamma',L'S'}\ ,
        \label{eq:wf}
\end{equation}
where the asymptotic wave functions $\Psi^{\lambda,JJ_z}_{\gamma,LS}$ with $\lambda=F,G$
are defined as
\begin{equation}
  \Psi^{\lambda,JJ_z}_{\gamma,LS} = { D_\gamma\over \sqrt{4}} \sum_{p=1}^{4}
    \Omega^{JJ_z}_{\gamma LS,p} { {\cal F}^\lambda_L(q_\gamma;y_{p}) \over
      q_\gamma y_{p} }\ ,   \label{eq:awf}
\end{equation}
and the superscript $\lambda=F$ is to denote the regular
radial functions introduced earlier in Eq.~(\ref{eq:pw1}), and $\lambda=G$
is to denote the irregular Coulomb or spherical Bessel
functions, namely
\begin{equation}
 \gamma=1\!:\, {\cal F}^{\, G}_L(x)= \widetilde G_L(\eta_1,x) \ ;
 \gamma=2\!:\, {\cal F}^{\, G}_L(x)= -x \widetilde y_L(x)\ . \label{eq:g}
\end{equation}
The tilde over $G_L$ and $y_L$ indicates that they have been multiplied by short-range
cutoffs in order to remove the singularity at the origin.  Thus ${\cal F}_L^{\, G}$ is
well-behaved in all space.  The normalization factor $D_\gamma$,
\begin{equation}
   D_\gamma=\sqrt{2\, \mu_\gamma q_\gamma\over \kappa^3}   \label{eq:d}
\end{equation}
and $\kappa=\sqrt{3/2}$, is introduced for convenience---$\kappa$ is a numerical factor relating
the inter-cluster separation ${\bf y}_p$ to the Jacobi variable ${\bf x}_{1p}$,
{\it i.e.}~${\bf x}_{1p}=\kappa \, {\bf y}_p$ (see Eq.~(\ref{eq:JcbV})
below). 

The wave functions $\Psi^{C,JJ_z}_{\gamma,LS}$ vanish in the asymptotic
region, and describe the dynamics of the interacting nucleons when they
are close to each other, while the $R^{\gamma \gamma',J}_{LS,L'S'}$'s are 
the $R$-matrix elements.  The latter, as well as the coefficients entering
the expansion of $\Psi^{C,JJ_z}_{\gamma,LS}$ in terms of hyperspherical-harmonics
functions, are determined via the Kohn variational principle
\begin{equation}
  \Bigl[R^{\gamma\gamma', J}_{LS,L'S'}\Bigr] = 
    R^{\gamma' \gamma, J}_{L'S',LS}-
    \langle\Psi^{JJ_z}_{\gamma,LS}|H-E|\Psi^{JJ_z}_{\gamma',L'S'}\rangle \ , \\
\label{eq:kohn}
\end{equation}
as discussed in Sec.~\ref{sec:whh}.

The next step consists in relating the $R$- to the $T$-matrix elements.  To this
end, it is convenient to simplify the notation by dropping the superscripts $JJ_z$
and by introducing a single label $\alpha$ to denote the channel quantum numbers $LS$,
so that the wave functions in Eq.~(\ref{eq:wf}) corresponding to $\gamma=1$ and 2
are written as
\begin{eqnarray}
\!\!\! \!\!\!    \Psi_{1,\alpha}\!\!\!&=&\!\!\!\Psi^C_{1,\alpha}\!+\!\Psi^F_{1,\alpha}\!
      + \!\sum_{\alpha'}\! R^{11}_{\alpha,\alpha'} \!\Psi^G_{1,\alpha'}
      + \!\sum_{\alpha'}\! R^{12}_{\alpha,\alpha'} \! \Psi^G_{2,\alpha'}\ ,  \label{eq:wf1}\\
\!\!\!\!\!\!     \Psi_{2,\alpha}\!\!\!&=&\!\!\!\Psi^C_{2,\alpha}\! +\!\Psi^F_{2,\alpha}
      + \!\sum_{\alpha'}\! R^{21}_{\alpha,\alpha'}\! \Psi^G_{1,\alpha'}\!
      + \!\sum_{\alpha'}\! R^{22}_{\alpha,\alpha'}\! \Psi^G_{2,\alpha'}\ .  \label{eq:wf2}
\end{eqnarray}
From these we form the linear combination
\begin{equation}
    \Psi=\sum_{\alpha'} \left(U_{\alpha,\alpha'} \Psi_{1,\alpha'}+ V_{\alpha,\alpha'} \Psi_{2,\alpha'}\right)\ ,
\end{equation}
where the matrices $U$ and $V$ are determined below.  Inserting the expressions
above for $ \Psi_{\gamma,\alpha}$ and rearranging terms lead to
\begin{widetext}
\begin{eqnarray}
    \Psi&=& \Psi^C + 
     \sum_{\alpha'} \Bigl[U-i\, (U R^{11}+V R^{21}) \Bigr]_{\alpha,\alpha'}
     \Psi^F_{1,\alpha'}
         + \sum_{\alpha'} \left( U R^{11}+V R^{21}\right)_{\alpha,\alpha'}
      \Bigl(\Psi^G_{1,\alpha'}+i\, \Psi^F_{1,\alpha'}\Bigr) 
    \nonumber\\ 
&+&\sum_{\alpha'} \Bigl[V-i\, (U R^{12}+V R^{22})\Bigr]_{\alpha,\alpha'} \Psi^F_{2,\alpha'}
    + \sum_{\alpha'} \left( U R^{12}+V R^{22}\right)_{\alpha,\alpha'}
      \Bigl(\Psi^G_{2,\alpha'}+i\Psi^F_{2,\alpha'}\Bigr)\ , 
\label{eq:wfUV1}
\end{eqnarray}
\end{widetext}
where $\Psi^C$ is a combination of internal parts of no interest here.  We now require
$\Psi$ to consist, in the asymptotic region, of a plane wave in channel $\gamma$=2 (or $n$-$^3$He) and of 
a purely outgoing wave in channel $\gamma$=1 (or $p$-$^3$H).  These requirements
are satisfied by demanding that
\begin{eqnarray}
   U-i\, (U R^{11}+V R^{21})&=&0\ , \\
 V-i\, (U R^{12}+V R^{22})&=&I\ ,
   \label{eq:UV} 
\end{eqnarray}
where $I$ is the identity matrix.  Comparing the resulting $\Psi$ with the wave function
given in Eq.~(\ref{eq:dw}), specifically its component in channel $LSJ$, allows one
to express the $T$-matrix as
\begin{eqnarray}
   T^{21,J}_{LS,L'S'}&=& {D_1\over D_2 \, q_1} \left( U^J R^{11,J}+V^J
   R^{21,J}\right)_{LS,L'S'} \nonumber \\
&=& -i \,{D_1\over D_2\, q_1}\, U^J_{LS,L'S'}\ ,  \label{eq:t21}
\end{eqnarray}
where we have reinstated the $LSJ$ labels.  Finally the matrix $U$ is obtained by solving
the system in Eq.~(\ref{eq:UV}):
\begin{widetext}
\begin{equation}
   T^{21,J}_{LS,L'S'}= {D_1\over D_2 \, q_1}\Bigg[
   \Bigl[I-i\,R^{22,J}+R^{21,J}(I-i\,R^{11,J})^{-1}R^{12,J}\Bigr]^{-1}
\, R^{21,J}
   (I-i\,R^{11,J})^{-1}\Bigg]_{LS,L'S'} \ .  \label{eq:t21b}
\end{equation}
\end{widetext}
In fact, we compute the $R$-matrix elements at zero energy, {\it i.e.}~in the limit $q_2 \rightarrow 0$,
and define
\begin{eqnarray}
  &&\overline R^{12,J}_{LS,L'S'}= {R^{12,J}_{LS,L'S'} \over q_2^{L'+1/2}}\ ,\qquad
   \overline R^{21,J}_{LS,L'S'}= {R^{21,J}_{LS,L'S'} \over q_2^{L+1/2}}\ , \nonumber \\
  &&\qquad\qquad \overline R^{22,J}_{LS,L'S'}= {R^{22,J}_{LS,L'S'} \over q_2^{L+L'+1}}\ ,   \label{eq:rbar}
\end{eqnarray}
and it can be shown that the $\overline R$-matrix elements are finite in this limit.  In particular,
we note that the factor $q_2^{L}$ follows from the small argument expansion of the spherical Bessel
function $j_L$ in $\Psi^{F,JJ_z}_{\gamma=2,LS}$, while the extra $q_2^{1/2}$ is due to the normalization $D_2$.  At zero energy, we have
\begin{equation}
 \Bigl[I-i\,R^{22,J}+R^{21,J}(I-i\,R^{11,J})^{-1}R^{12,J}\Bigr] \rightarrow I \ ,
\end{equation}
since $R^{22,J}$ and the product $R^{21,J}R^{12,J}$ are proportional to $q_2$ or
higher powers of $q_2$.  Furthermore, the relevant $T$-matrix elements entering
the expression for the asymmetry $A_z$ are those with quantum number $L=0$
in channel $\gamma=2$, and hence
\begin{equation}
   T^{21,J}_{0J,L'S'}= {1\over \sqrt{q_1}} 
    \sum_{L''S''}\overline R^{21,J}_{0J,L''S''} \left(I-i\, R^{11,J}\right)^{-1}_{L''S'',L'S'} \ ,
  \label{eq:t21c}
\end{equation}
with $J=0,1$.  Note that we have neglected the difference in the $n$-$^3$He and $p$-$^3$H reduced
masses.

\section{The parity-violating potential}
\label{sec:pv}

Two different models of the PV weak-interaction potentials are
adopted in the calculations reported below.  One is the model
developed thirty years ago by Desplanques
{\it et al.}~\cite{Desplanques80} (and known as DDH): it
is parametrized in terms of $\pi$-, $\rho$-, and $\omega$-meson
exchanges, and involves in practice six weak pion and
vector-meson coupling constants to the nucleon~\cite{Note1}.  These
were estimated within a quark model approach incorporating
symmetry arguments and current algebra
requirements~\cite{Desplanques80,Holstein81}.  Due to
the inherent limitations of such an analysis, however,
the coupling constants determined in this way have rather wide
ranges of allowed values.

The other model for the PV potential considered in the present
work is that formulated by Zhu {\it et al.}~\cite{Zhu05} in 2005,
and reduced to its minimal form by Girlanda~\cite{Girlanda08} in 2008,
within an effective-field-theory (EFT) approach in which
only nucleon degrees of freedom are retained explicitly.
At lowest order $Q/\Lambda_\chi$, where $Q$ is the small
momentum scale characterizing the low-energy PV process and $\Lambda_\chi
\simeq 1$ GeV is the scale of chiral symmetry breaking,
it is parametrized by a set of five contact four-nucleon terms.
\begin{widetext}
\begin{center}
\begin{table}[btt]
\begin{tabular}{c|c|c|c|c|c}
\hline
\hline
$n$      & $c^{\rm DDH}_n$  & $f^{\rm DDH}_n(r)$ 
         & $c^{\rm EFT}_n$  & $f^{\rm EFT}_n(r)$ & $O^{(n)}_{ij}$ \\
\tableline
1        & $+\frac{g_\pi \, h^1_\pi}{2\sqrt{2} \, m}$ & $f_\pi(r)$  &
$\frac{2\, \mu^2}{\Lambda_\chi^3}\, C_6$  &  
$f_\mu(r)$ & $
({\bm \tau}_i\times {\bm \tau}_j)_z \,({\bm \sigma}_i+{\bm \sigma}_j)\cdot
{\bf X}^{(1)}_{ij,-}$ \\ 
2        & $-\frac{g_\rho \, h^0_\rho}{ m}$  & $f_\rho(r)$  & 0  & 0  & 
${\bm \tau}_i \cdot {\bm \tau}_j \,({\bm \sigma}_i-{\bm \sigma}_j)\cdot
{\bf X}^{(2)}_{ij,+} $          \\
3        & $-\frac{g_\rho \, h^0_\rho(1+\kappa_\rho)}{ m}$ & $f_\rho(r)$  & 0 & 0  & 
${\bm \tau}_i \cdot {\bm \tau}_j \,({\bm \sigma}_i\times{\bm \sigma}_j)\cdot {\bf X}^{(3)}_{ij,-} $      \\
4        & $-\frac{g_\rho \, h^1_\rho}{2\, m}$ & $f_\rho(r)$  & $\frac{ \mu^2}{\Lambda_\chi^3}\, (C_2+C_4)$  &  $f_\mu(r)$ & $
({\bm \tau}_i + {\bm \tau}_j)_z\,({\bm \sigma}_i-{\bm \sigma}_j)\cdot 
{\bf X}^{(4)}_{ij,+} $          \\
5        & $-\frac{g_\rho \, h^1_\rho(1+\kappa_\rho)}{2\, m}$ & $f_\rho(r)$ & 0 & 0 &
$({\bm \tau}_i + {\bm \tau}_j)_z \,({\bm \sigma}_i\times{\bm \sigma}_j)\cdot 
{\bf X}^{(5)}_{ij,-} $   \\
6        & $-\frac{g_\rho \, h^2_\rho}{2 \sqrt{6}\, m}$ & $f_\rho(r)$  & $-\frac{2\, \mu^2}{\Lambda_\chi^3}\, 
C_5$  & $f_\mu(r)$  &
$ (3\, \tau_{i,z} \tau_{j,z}-{\bm \tau}_i \cdot {\bm \tau}_j)
 \,({\bm \sigma}_i-{\bm \sigma}_j)\cdot 
{\bf X}^{(6)}_{ij,+} $       \\
7        & $-\frac{g_\rho \, h^2_\rho(1+\kappa_\rho)}{2\sqrt{6}\, m}$ & $f_\rho(r)$ & 0 & 0 & $ (3\,  \tau_{i,z} \tau_{j,z}-{\bm \tau}_i 
\cdot {\bm \tau}_j) \,({\bm \sigma}_i\times {\bm \sigma}_j)\cdot 
{\bf X}^{(7)}_{ij,-} $  \\
8        & $-\frac{g_\omega \, h^0_\omega}{ m}$  & $f_\omega(r)$  & $\frac{2\, \mu^2}{\Lambda_\chi^3} 
\, C_1$   & $f_\mu(r)$ & $({\bm \sigma}_i-{\bm \sigma}_j)\cdot {\bf X}^{(8)}_{ij,+} $          \\
9        & $-\frac{g_\omega \, h^0_\omega (1+\kappa_\omega)}{ m}$ & $f_\omega(r)$ &  $\frac{2\, \mu^2}{\Lambda_\chi^3} \,\tilde{C}_1$   & $f_\mu(r)$ & $ ({\bm \sigma}_i\times {\bm \sigma}_j)\cdot 
{\bf X}^{(9)}_{ij,-} $   \\
10       & $-\frac{g_\omega \, h^1_\omega}{2\, m}$ & $f_\omega(r)$  &  0 & 0 & $
({\bm \tau}_i + {\bm \tau}_j)_z\, ({\bm \sigma}_i-{\bm \sigma}_j)\cdot 
{\bf X}^{(10)}_{ij,+} $           \\
11       & $-\frac{g_\omega \, h^1_\omega(1+\kappa_\omega)}{2\, m}$ & $f_\omega(r)$  &  0 & 0 & $ ({\bm \tau}_i + {\bm \tau}_j)_z\, ({\bm \sigma}_i\times {\bm \sigma}_j)\cdot 
{\bf X}^{(11)}_{ij,-} $  \\
12       & $-\frac{g_\omega h^1_\omega -g_\rho h^1_\rho}{2\, m}$ & $f_\rho(r)$
& 0 &  0 & 
$ ({\bm \tau}_i -{\bm \tau}_j)_z \,({\bm \sigma}_i+{\bm \sigma}_j)\cdot 
{\bf X}^{(12)}_{ij,+} $           \\
\hline
\hline
\end{tabular}
\caption{Components of the DDH and EFT models for the
parity-violating potential.  The vector operators ${\bf X}^{(n)}_{ij,\mp}$ and
functions $f_x(r)$, $x=\pi,\, \rho,\, \omega,\, \mu$, are defined in
Eqs.~(\protect\ref{eq:x+})--(\protect\ref{eq:x-}) and
Eqs.~(\protect\ref{eq:fx})--(\ref{eq:fmu}), respectively.
Only 5 operators and low-energy
constants enter the pionless EFT interaction at the leading order, and in
this paper they have been chosen to correspond to the rows 1, 4, 6, 8 and 9.}
\label{tb:tab1}
\end{table}
\end{center}
\end{widetext}
The DDH and EFT PV two-nucleon potentials are conveniently written as
\begin{equation}
 v^\alpha_{ij}=\sum_{n=1}^{12} c^\alpha_n \, O^{(n)}_{ij} \ ,
  \qquad \alpha={\rm DDH} \,\, {\rm or}\,\, {\rm EFT} \ ,
\end{equation}
where the parameters $c^\alpha_n$ and operators $O^{(n)}_{ij}$, $n=1,\dots, 12$,
are listed in Table~\ref{tb:tab1}.  In this table the vector operators
${\bf X}^{(n)}_{ij,\pm}$ are defined as
\begin{eqnarray}
 {\bf X}^{(n)}_{ij,+} &\equiv& \left[ {\bf p}_{ij} \, ,\, f_n(r_{ij}) \right]_+ \ , 
\label{eq:x+} \\
 {\bf X}^{(n)}_{ij,-} &\equiv& {\rm i}\, \left[ {\bf p}_{ij} \, ,\, f_n(r_{ij}) \right]_- \ ,
\label{eq:x-}
\end{eqnarray}
where $\left[ \dots \, , \, \dots \right]_{\mp}$ denotes the commutator ($-$)
or anticommutator ($+$), and ${\bf p}_{ij}$ is the relative momentum operator, 
${\bf p}_{ij} \equiv ({\bf p}_i-{\bf p}_j)/2$.  In the DDH model,
the functions $f_x(r)$, $x=\pi, \rho$ and $\omega$, are Yukawa functions,
suitably modified by the inclusion of monopole form factors,
\begin{equation}
 f_x(r) =\frac{1}{4\pi\, r} \left\{
 {\rm e}^{-m_x r} -{\rm e}^{-\Lambda_x r}\left[1+
 \frac{\Lambda_x r }{2}\left(1-\frac{m_x^2}
 {\Lambda_x^2}\right)\right] \right\} \ .
\label{eq:fx}
\end{equation}
In the EFT model, however, the short-distance behavior is
described by a single function $f_\mu(r)$, which is
itself taken as a Yukawa function with mass parameter $\mu$,
\begin{equation}
 f_\mu(r)=\frac{1}{4\pi\, r}{\rm e}^{-\mu r} \ ,
\label{eq:fmu}
\end{equation}
with $\mu \simeq m_\pi$ as appropriate in
the present formulation, in which pion degrees of
freedom are integrated out.

In the potential $v^{\rm DDH}_{ij}$, the strong-interaction coupling constants
of the $\pi$-, $\rho$-, and \hbox{$\omega$-meson} to the nucleon are denoted
as $g_\pi,\, g_\rho,\, \kappa_\rho,\, g_\omega,\, \kappa_\omega$,
while the weak-interaction ones as $h^1_\pi,\, h^0_\rho,\, h^1_\rho,\,
 h^2_\rho,\, h^0_\omega,\, h^1_\omega$, where
the superscripts 0, 1, and 2 specify the isoscalar, isovector,
and isotensor content of the corresponding interaction components.
In the EFT model, the five low-energy constants $C_1, \tilde{C}_1,
C_2+C_4 , C_5$ and $C_6$
completely characterize $v^{\rm EFT}_{ij}$, to lowest order
$Q/\Lambda_\chi$.

\begin{table}[bth]
\begin{tabular}{c|c|c|c|c|c|c}
\hline
\hline
    & $g^2_\alpha/4\pi$  & $\kappa_\alpha$  & $10^7\times h_\alpha^0$  & $10^7\times h_\alpha^1$
&  $10^7\times h_\alpha^2 $ & $\Lambda_\alpha$ (GeV/c)  \\
\hline
$\pi$          & 13.2 &     &        &   4.56 &       & 1.72 \\
$\rho$       & 0.840 & 6.1  & --16.4 & --2.77 & --13.7 & 1.31 \\
$\omega$ & 20.0  & 0.0 & 3.23  & 1.94 &       & 1.50 \\
\hline
\hline
\end{tabular}
\caption{Values used for the strong- and weak-interaction coupling
constants and short-range cutoff parameters of the $\pi$-, $\rho$-,
and $\omega$-meson in the DDH potential.}
\label{tb:tabpv}
\end{table}
The values for the coupling constants and short-range cutoffs in the DDH model are
listed in Table~\ref{tb:tabpv}, while the mass $\mu$ in the EFT model is taken to be
$m_\pi$.   These values for coupling constants and cutoffs were also
used in the DDH-based calculations of PV two-nucleon observables in
Refs.~\cite{Carlson02,Schiavilla04} and neutron spin rotation in
$\vec{n}\, d$ scattering~\cite{Schiavilla08}.  In particular, we note
that the linear combination of $\rho$- and $\omega$-meson
weak coupling constants corresponding to $pp$ states has
been taken from an earlier analysis of $\vec{p}\, p$ elastic
scattering experiments~\cite{Carlson02}.  The remaining couplings are
the ``best value'' estimates, suggested in Ref.~\cite{Desplanques80}.

In the analysis of the $a_z$ observable to follow,
we will report results for the coefficients $I^{\rm DDH}_n$ and
$I^{\rm EFT}_n$ in the expansion
\begin{equation}
a_z =
\sum_{n=1}^{12} c^\alpha_n \, I^\alpha_n \ . 
\label{eq:in}
\end{equation}
Thus we will not need to consider specific values (or
range of values) for the strength parameters $c^\alpha_n$.
However, the $I^\alpha_n$ depend on the masses (and short-range
cutoffs $\Lambda_x$ for the DDH model) occurring in the Yukawa
functions.  Note that the coefficients $C^i_\alpha$ entering
Eq.~(\ref{eq:cddh}) are obtained from the $I^{\rm DDH}_n$'s and
$c^{\rm DDH}_n$'s listed in Table~\ref{tb:tab1} via
\begin{eqnarray}
C^1_\pi &=& +\frac{g_\pi }{2\sqrt{2} \, m} I^{\rm DDH}_1\ , \nonumber \\
C^0_\rho &=& -\frac{g_\rho}{ m} I^{\rm DDH}_2 
             -\frac{g_\rho \,(1+\kappa_\rho)}{ m} I^{\rm DDH}_3\ , \nonumber \\
C^1_\rho &=&  -\frac{g_\rho }{2\, m} I^{\rm DDH}_4
                -\frac{g_\rho (1+\kappa_\rho)}{2\, m} I^{\rm DDH}_5
                +\frac{g_\rho}{2\, m} I^{\rm DDH}_{12}\ , \nonumber  \\
C^2_\rho &=&  -\frac{g_\rho}{2 \sqrt{6}\, m} I^{\rm DDH}_6
              -\frac{g_\rho (1+\kappa_\rho)}{2\sqrt{6}\, m}  I^{\rm DDH}_7\ ,  \\
C^0_\omega &=&  -\frac{g_\omega}{ m} I^{\rm DDH}_8
                -\frac{g_\omega (1+\kappa_\omega)}{ m} I^{\rm DDH}_9\ , \nonumber \\
C^1_\omega &=&  -\frac{g_\omega }{2\, m} I^{\rm DDH}_{10} 
                -\frac{g_\omega (1+\kappa_\omega)}{2\, m}I^{\rm DDH}_{11} 
                -\frac{g_\omega }{2\, m}I^{\rm DDH}_{12}\ . \nonumber 
\end{eqnarray}

\section{The HH wave functions}
\label{sec:whh}

The \lq\lq internal\rq\rq wave function $\Psi^{C,JJ_z}_{\gamma,LS}$, see Eq.~(\ref{eq:wf}), is
expanded in the HH basis.  For four equal mass particles, a suitable choice for
the Jacobi vectors is
\begin{eqnarray}
   {\bf x}_{1p}& = & \sqrt{\frac{3}{2}} 
    \left ({\bf r}_l - \frac{ {\bf r}_i+{\bf r}_j +{\bf r}_k}{3} \right )\ , \nonumber\\
  {\bf x}_{2p} & = & \sqrt{\frac{4}{3}}
    \left ({\bf r}_k-  \frac{ {\bf r}_i+{\bf r}_j}{2} \right )\ , \label{eq:JcbV}\\
   {\bf x}_{3p} & =& {\bf r}_j-{\bf r}_i\ , \nonumber
\end{eqnarray}
where $p$ specifies a given permutation corresponding to the ordering $(ijkl)$.
By definition, the permutation $p=1$ is chosen to correspond to (1234).

For the given Jacobi vectors, the hyperspherical coordinates include
the so-called hyperradius $\rho$, defined by
\begin{equation}
   \rho=\sqrt{x_{1p}^2+x_{2p}^2+x_{3p}^2}\quad ({\rm independent\
    of\ }p)\ ,
    \label{eq:rho}
\end{equation}
and a set of angular variables which in the Zernike and
Brinkman~\cite{zerni,F83} representation are (i) the polar angles $\hat
{\bf x}_{ip}\equiv (\theta_{ip},\phi_{ip})$ of each Jacobi vector, and (ii) the
two additional ``hyperspherical'' angles $\phi_{2p}$ and $\phi_{3p}$,
defined as
\begin{equation}
    \cos\phi_{2p} = \frac{ x_{2p} }{\sqrt{x_{1p}^2+x_{2p}^2}}\ ,
    \quad
    \cos\phi_{3p} = \frac{ x_{3p} }{\sqrt{x_{1p}^2+x_{2p}^2+x_{3p}^2}}\ ,
     \label{eq:phi}
\end{equation}
where $x_{jp}$ is the magnitude of the Jacobi vector ${\bf x}_{jp}$. The set of angular
variables $\hat {\bf x}_{1p}, \hat {\bf x}_{2p}, \hat {\bf x}_{3p}, \phi_{2p}$, and $\phi_{3p}$ is
denoted  hereafter as $\Omega_p$.  A generic HH function reads
\begin{widetext}
\begin{eqnarray}
{\cal H}^{K\Lambda M}_{\ell_1 \ell_2\ell_3 L_2 n_2 n_3}(\Omega_p) &=&
   {\cal N}^{\ell_1 \ell_2 \ell_3}_{ n_2 n_3} 
      \left [ \Bigl [ Y_{\ell_1}(\hat {\bf x}_{1p}) \otimes
    Y_{\ell_2}(\hat{\bf x}_{2p}) \Bigr ]_{L_2} \otimes Y_{\ell_3}(\hat {\bf x}_{3p}) \right
    ]_{\Lambda M}
   (\sin\phi_{2p})^{\ell_1 }    (\cos\phi_{2p})^{\ell_2}
(\sin\phi_{3p})^{\ell_1+\ell_2+2n_2} \nonumber \\
&\times&
(\cos\phi_{3p})^{\ell_3}\,
      P^{\ell_1+1/2\, , \,\ell_2+1/2}_{n_2}(\cos2\phi_{2p})
      P^{\ell_1+\ell_2+2\, n_2+2\, , \,\ell_3+1/2}_{n_3}(\cos2\phi_{3p})\ ,
      \label{eq:hh4P}
\end{eqnarray}
\end{widetext}
where $P^{a,b}_n$ are Jacobi polynomials, and the coefficients
${\cal N}^{\ell_1 \ell_2 \ell_3}_{ n_2  n_3}$ are normalization factors.
The quantity $K=\ell_1+\ell_2+\ell_3+2\,(n_2+n_3)$ is the so-called
grand angular quantum number.  The HH functions are the eigenfunctions
of the hyperangular part of the kinetic energy operator.  Another important
property is that $\rho^K \,  {\cal  H}^{K \Lambda M}_{\ell_1 \ell_2 \ell_3  L_2 n_2
n_3}(\Omega_p)$ are homogeneous polynomials of the particle coordinates of
degree $K$.

A set of antisymmetrized hyperangular-spin-isospin states of 
grand angular quantum number $K$, total orbital angular momentum $\Lambda$,
total spin $\Sigma$, and total isospin $T$  (for the given values of
total angular momentum $J$ and parity $\pi$) can be constructed as follows:
\begin{equation}
  \Psi_{\mu}^{K\Lambda\Sigma T} = \sum_{p=1}^{12}
  \Phi_\mu^{K\Lambda\Sigma T}(ijkl)\ ,
  \label{eq:PSI}
\end{equation}
where the sum is over the $12$ even permutations $p\equiv ijkl$, and
\begin{widetext}
\begin{equation}
  \Phi^{K\Lambda\Sigma T}_{\mu}(ijkl)
   = \biggl [
   {\cal H}^{K \Lambda M}_{\ell_1\ell_2\ell_3 L_2  n_2 n_3}(\Omega_p) \otimes
      \biggl [\Bigl[\bigl[ \chi_i \otimes \chi_j \bigr]_{S_a}\otimes
      \chi_k\Bigr]_{S_b} \otimes \chi_l  \biggr]_{\Sigma} \biggr ]_{JJ_z}
   \biggl [\Bigl[\bigl[ \xi_i \otimes \xi_j \bigr]_{T_a}
      \otimes \xi_k\Bigr]_{T_b} \otimes \xi_l  \biggr]_{TT_z}\ .
     \label{eq:PHI}
\end{equation}
\end{widetext}
Here, $\chi_i$ ($\xi_i$) denotes the spin 
(isospin) state of particle $i$.  The total orbital angular momentum $\Lambda$ of
the HH function is coupled to the total spin $\Sigma$ to give the total angular
momentum $JJ_z$, whereas the parity $\pi$ is $(-1)^{\ell_1+\ell_2+\ell_3}$.  The
quantum number $T$ specifies the total isospin of the state, and
$\mu$ labels the possible choices of hyperangular, spin and isospin
quantum numbers, namely
\begin{equation}
   \mu \equiv \{ \ell_1,\ell_2,\ell_3, L_2 ,n_2, n_3, S_a,S_b, T_a,T_b
   \}\ ,\label{eq:mu}
\end{equation}
compatible with the given values of $K$, $\Lambda$, $\Sigma$,
$T$, $J$, and $\pi$.  Another important classification scheme for
the states is to group them in ``channels'': states
belonging to the same channel have the same values of angular
($\ell_1,\ell_2,\ell_3, L_2 ,\Lambda$), spin ($S_a,S_b,\Sigma$),
and isospin ($T_a,T_b,T$) quantum
numbers, but different values of $n_2$ and $n_3$.

Each state  $\Psi^{K\Lambda\Sigma T}_\mu$ entering
the expansion of the four-nucleon wave function must
be antisymmetric under the exchange of any pair of
particles.  Consequently, it is necessary to consider states
such that
\begin{equation}
    \Phi^{K\Lambda\Sigma T}_\mu(ijkl)= 
    -\Phi^{K\Lambda\Sigma T}_\mu(jikl)\ ,
     \label{eq:exij}
\end{equation}
which is fulfilled when the condition
\begin{equation} 
    \ell_3+S_a+T_a = {\rm odd}\ , \label{eq:lsa}
\end{equation}
is satisfied.

The number $M_{K\Lambda\Sigma T}$ of  antisymmetrized functions
$\Psi^{K\Lambda\Sigma T}_\mu$ having given values of $K$, $\Lambda$, $\Sigma$,
and $T$, but different combinations of quantum numbers $\mu$---see Eq.(\ref{eq:mu})---is
in general very large.  In addition to the degeneracy of the HH basis, the four
spins (isospins) can be coupled in different ways to total $\Sigma$ ($T$).  However, many
of the states $\Psi^{K\Lambda\Sigma T}_\mu$, with $\mu$ ranging from 1
to $M_{K\Lambda\Sigma T}$, are linearly dependent.  In the expansion of $\Psi^{C,JJ_z}_{\gamma,LS}$,
it is necessary to include only the subset of linearly independent states, whose
number is fortunately significantly smaller than $M_{K\Lambda\Sigma T}$.

The internal part of the  wave function can be finally written as
\begin{equation}\label{eq:PSI3}
  \Psi^{C,JJ_z}_{\gamma,LS}= \sum_{K\Lambda\Sigma T}\sum_{\mu} 
    u^{\gamma,LS}_{K\Lambda\Sigma T\mu}(\rho)
    \Psi_{\mu}^{K\Lambda\Sigma T}\ ,
\end{equation}
where the sum is restricted only to the linearly independent states. 
We have found it convenient to expand the ``hyperradial'' functions
$u^{\gamma,LS}_{K\Lambda\Sigma T\mu}(\rho)$ in a 
complete set of functions, namely
\begin{equation}
     u^{\gamma,LS}_{K\Lambda\Sigma T\mu}(\rho)=\sum_{m=0}^{M-1} 
      c^{\gamma,LS}_{K\Lambda\Sigma T\mu,m} \; g_m(\rho)
      \ ,     \label{eq:fllag}
\end{equation}
and have chosen 
\begin{equation}
   g_m(\rho)= 
     \sqrt{\frac{m!}{(m+8)!}}\,\beta^{9/2}\, 
     L^{(8)}_m(\beta\rho)\,\,{\rm e}^{-\beta \rho/2} \ ,
      \label{eq:fllag2}
\end{equation}
where $L^{(8)}_l(\beta\rho)$ are Laguerre polynomials~\cite{abra}.

The $c$ coefficients of the expansion~(\ref{eq:fllag}) and the R-matrix
elements of Eq.~(\ref{eq:wf}) are determined 
variationally via the Kohn variational principle. This principle states that
the functional $\Bigl[R^{\gamma\gamma', J}_{LS,L'S'}\Bigr]$ defined in
Eq.~(\ref{eq:kohn}) is stationary with respect to variations in the
$R^{\gamma\gamma',J}_{LS,L^\prime S^\prime}$ and
$c^{\gamma,LS}_{K\Lambda\Sigma T\mu,m}$.   
By applying this principle, a linear set of equations for 
$R^{\gamma\gamma',J}_{LS,L^\prime S^\prime}$ and
$c^{\gamma,LS}_{K\Lambda\Sigma T\mu,m}$  is obtained~\cite{Kievsky08}, then 
solved using the Lanczos algorithm.  The other parameter entering the
expansion is the (non linear) parameter $\beta$ (see Eq.~(\ref{eq:fllag2})), 
used to describe the hyperradial functions $u^{\gamma,LS}_{K\Lambda\Sigma
  T\mu}(\rho)$. We have checked that, once a sufficient number $M$ of
functions $g_m(\rho)$ are employed ($M\approx 20$), the results are practically 
independent on $\beta$. In the present work we 
have used $\beta=4$ fm$^{-1}$.

The application of the method has two main difficulties.  The first is the
accurate computation of the matrix elements of the Hamiltonian.  By exploiting
the properties of the HH functions, however, this task can be noticeably
simplified, as discussed in Refs.~\cite{Kievsky08,Viviani05}.  The second
difficulty is the slow convergence of the HH expansion.  This problem has been
overcome by dividing the set of states $\Psi_{\mu}^{K\Lambda\Sigma
  T}$ defined in Eq.~(\ref{eq:PSI}) (in the following referred to simply as
``HH states'') in \textit{classes}, depending on the value of
$\mathcal{L} = \ell_1 + \ell_2 + \ell_3$, total isospin $T$, and $n_2$ and 
$n_3$. In the present paper, we have considered four different classes. 
Since for $n$-$^3$He scattering the asymptotic states do not have a definite
total isospin (they are a superposition of $T=0$ and $T=1$ components),
it is mandatory to include HH states with both $T=0$ and $1$. The
contribution of $T=2$ states is expected to be tiny and consequently
they have been ignored in the present paper.

Following Refs.~\cite{Viviani05,Fisher06}, in the first class we have included
the $n_2=0$ HH states belonging to some special channels, for which the
convergence has been found to be critical.  The radial part of these HH states
depends only on $\cos\phi_{3p}=r_{ij}/\rho$ and thus they take into account 
two-body correlations. The $n_2>0$ HH states
belonging to the same channels are included in the second class, together
with those having $\mathcal{L}\le 2$.  The other classes are then
defined simply by grouping HH states belonging to channels with an increasing
value of $\mathcal{L}$. In particular, for the construction of the positive
(negative) parity \lq\lq internal\rq\rq wave function
$\Psi^{C,JJ_z}_{\gamma,LS}$, classes 3 and 4 include all HH states with
$\mathcal{L}=4$ and $6$ ($\mathcal{L}=3$ and $5$), respectively. The
convergence of these last two classes is less critical, and consequently, only
HH states with lower values of grand angular quantum number $K$ need be
considered. Moreover, the  convergence with $\mathcal{L}$ is quite fast.  In
particular, we have found that, at the energy considered, the contribution of
HH states with $\mathcal{L}>6$ can be neglected.

The calculation is performed including in the expansion all 
HH states belonging to classes $i=1,\ldots,4$ with grand angular quantum
number $K\le K_i$, where $K_1,\ldots,K_4$ are a set of nonnegative integers.
The convergence of a quantity of interest (for example, the
phase-shifts, or the coefficient $a_z$ defining the PV asymmetry) is 
then studied by increasing the values of $K_i$. A more complete study of the
convergence will be presented elsewhere~\cite{Viviani10}.

To exhibit the convergence pattern, we report in Table~\ref{tb:conv1}
the calculated $n$-$^3$He scattering lengths.  As is evident from Eq.~(\ref{eq:dw}),
they are defined as 
\begin{equation}
a_J=-\lim_{q_2\rightarrow0} T^{22,J}_{0J,\,0J} \ ,
\end{equation}
with both incoming and outgoing $n$-$^3$He clusters in relative S-wave. Note that in
general this scattering length is complex, since the channel
$p$-$^3$H is always open, and therefore the unitarity condition imposes
that ${\rm Im}\, a_J <0$, since the total cross section
is proportional to $\sum_{J=0,1} (2J+1)\,{\rm Im}\, T^{22,J}_{0J,\,0J}$.
The results obtained for the singlet ($J=0$) and triplet ($J=1$)
scattering lengths are reported in Table~\ref{tb:conv1}, for all four
potential models used in this work.  The calculated $n$-$^3$He scattering
lengths are compared with experimental values and the results of other
calculations available in the literature.

Inspection of the table shows that the convergence for the triplet scattering
length is very good, 
and that there is reasonable agreement with available experimental values, and the results
of other calculations, in particular those of the AGS method.  In the case of
the singlet scattering length, the situation is more
delicate, since in the channel $J^\pi=0^+$ the $n$-$^3$He interaction is
attractive and the wave function must be orthogonal to the $^4$He bound
state.  Consequently, the convergence is more problematic, in particular for
the N3LO/N2LO interaction model.  In the row labeled
``EXT'', we have reported the extrapolated values for this quantity obtained
by analyzing the convergence pattern.  For the AV18, N3LO, and AV18/UIX
interaction models we observe reasonable agreement with the results of other
calculations and the experimental data.  The N3LO/N2LO values are significantly different
from those obtained with the other interaction
models, which is presumably related to the slow convergence observed in this
case.  A complete study of the $n$-$^3$He scattering lengths is in
progress~\cite{Viviani10}.

\begin{widetext}
\begin{center}
\begin{table}[h]
\begin{tabular}{c@{$\quad$}|c@{$\quad$}|c@{$\quad\ $}|c@{$\quad\ $}
                @{$\ $} |c@{$\ $}|c@{$\ $}  
                @{$\ $} |c@{$\ $}|c@{$\ $} }
\hline
\hline
  \multicolumn{8}{c}{Triplet scattering length $a_1$ (fm)} \\
\hline 
$K_1$ & $K_2$ & $K_3$ & $K_4$ & 
     AV18  & N3LO & AV18/UIX & N3LO/N2LO \\
\hline
 28 & 28 & 20 & 20 & $3.56- i\, 0.0078$ & $3.47-i\, 0.0047$ &
                       $3.39-i\, 0.0059$ & $3.37-i\, 0.0042$ \\
 30 & 30 & 22 & 22 & $3.56-i\, 0.0077$ & $3.46-i\, 0.0048$ &
                       $3.39-i\, 0.0059$ & $3.37-i\, 0.0042$ \\
\hline
 \multicolumn{4}{l|}{RGM~\protect\cite{HH08}} 
        & $3.45-i\, 0.0066$ &   & $3.31-i\, 0.0051$ & \\
 \multicolumn{4}{l|}{FY~\protect\cite{Lazauskas09}}  
        & $3.43-i\, 0.0082$ & $3.56-i\, 0.0070$ &
                             $3.23-i\, 0.0054$ & \\
 \multicolumn{4}{l|}{AGS~\protect\cite{Deltuvapv}} &  $3.51-i\,0.0074$  & $3.47-i\, 0.0068$ 
                         & & \\
\hline
 \multicolumn{4}{l}{R-matrix~\protect\cite{HH08}} 
        & \multicolumn{4}{l}{$3.29\phantom{(6)}-i\, 0.0012\phantom{(2)}$}  \\
 \multicolumn{4}{l}{EXP~\protect\cite{Zimmer02}}    
        & \multicolumn{4}{l}{$3.28(5)-i\, 0.001(2)$}  \\
 \multicolumn{4}{l}{EXP~\protect\cite{Huffman04}}
        & \multicolumn{4}{l}{$3.36(1)\phantom{-i\, 0.001(2)}$}  \\
 \multicolumn{4}{l}{EXP~\protect\cite{Ketter06}}    
        & \multicolumn{4}{l}{$3.48(2)\phantom{-i\, 0.001(2)}$}  \\
\hline
\hline
\multicolumn{8}{c}{Singlet scattering length $a_0$ (fm)} \\
\hline 
$K_1$ & $K_2$ & $K_3$ & $K_4$ & 
     AV18  & N3LO & AV18/UIX & N3LO/N2LO \\
\hline
 48 & 44 & 30 & 22 & $7.34-i\, 6.27$ & $7.38-i\, 5.23$ &
                             $7.90-i\, 3.65$ & $4.45-i\, 9.02$ \\
 50 & 46 & 32 & 24 & $7.41-i\, 6.16$ & $7.40-i\, 5.20$ &
                             $7.90-i\, 3.59$ & $5.25-i\, 9.25$ \\
\hline 
\multicolumn{4}{l|}{EXT} & $7.69-i\, 5.70$ & $7.57-i\, 4.97$ &
                           $7.89-i\, 3.44$ & $6.02 -i\, 9.48 $ \\
\hline
 \multicolumn{4}{l|}{RGM~\protect\cite{HH08}} & $7.78-i\, 5.02$ & 
                         & $7.62-i\, 4.09$ & \\
 \multicolumn{4}{l|}{AGS~\protect\cite{Deltuvapv}} & $7.80-i\,4.97$  & $7.82-i\, 4.51$ 
                         & & \\
\hline
 \multicolumn{4}{l}{R-matrix~\protect\cite{HH08}} 
            & \multicolumn{4}{l}{$7.40\phantom{(6)}-i\, 4.449\phantom{(5)}$}  \\
 \multicolumn{4}{l}{EXP~\protect\cite{Zimmer02}}    
            & \multicolumn{4}{l}{$7.37(6)-i\, 4.448(5)$}  \\
 \multicolumn{4}{l}{EXP~\protect\cite{Huffman04}}    
            & \multicolumn{4}{l}{$7.46(2)\phantom{-i\, 4.448(5)}$}  \\
 \multicolumn{4}{l}{EXP~\protect\cite{Ketter06}}    
            & \multicolumn{4}{l}{$7.57(3)\phantom{-i\, 4.448(5)}$}  \\
\hline
\end{tabular}
\caption[Table]{\label{tb:conv1}
Convergence of the $n$-$^3$He singlet and triplet scattering lengths
corresponding to the inclusion, in the internal part of the wave function,
of four different classes in which the HH basis has been
subdivided.  For the singlet scattering length, the line labeled ``EXT''
reports the extrapolated values obtained by examining the
convergence pattern with increasing number of HH functions in the
expansion.  The calculated scattering lengths are
compared with results obtained using the Resonating Group Method (RGM),
Faddeev-Yakubovsky (FY) equations, Alt-Grassberger-Sandhas (AGS)
equations, as well as with results of R-matrix analyses.  The experimental
values are reported in the rows labeled ``EXP'' (the
imaginary parts are taken from Ref.~\protect\cite{HH08}).
}
\end{table}
\end{center}
\end{widetext}

Recently, there has been a new measurement~\cite{Huber09} for the quantity
$\Delta a'={\rm Re}(a_1-a_0)= -4.20(3)$ fm.  The calculated values of $\Delta
a'$ with the AV18, N3LO, AV18/UIX, and N3LO/N2LO models are
$-4.13$, $-4.11$, $-4.50$, and $-2.65$ fm, respectively.  Again the N3LO/N2LO value
stands out: it is off that obtained with the other interaction models and the measured value.

The convergence for the negative-parity states is similar to that discussed
above.  For the $0^-$ state, there is a close resonant state and the convergence
is slow as in the $0^+$ case.  For the $1^-$ state, the resonance is far and we
observe good convergence, as for the $1^+$ state.  Note, however, that in these
cases the N3LO/N2LO convergence pattern
is not different from that observed with the other models.

\section{Calculation}
\label{sec:calc}

There is a total of two (four) states with $J=0$ ($J=1$): one (two) with positive
parity having $LS=00$ ($LS=01, 21$) and one (two) with negative parity having
$LS=11$ ($LS=10,11$).  The $R$-matrix elements $R^{\gamma \gamma',0}_{LS,LS}$
with $LS=00$ or $LS=11$ for $J=0$, and
$R^{\gamma \gamma',1}_{LS,L'S'}$ with $LS,L'S'=01,21$ or $LS,L'S'=10,11$
for $J=1$, involving parity-conserving transitions induced by the strong
interactions are calculated with the HH method, as described in the previous section.
However, the $R$-matrix elements involving parity-violating (PV) transitions are
obtained in first-order perturbation theory as
\begin{equation}
 R^{\gamma \gamma',J}_{LS,L'S'}=- 
       \langle\Psi^{JJ_z} _{\gamma',L'S'}\mid v^{PV}\mid \Psi^{JJ_z}_{\gamma,LS}\rangle \ ,
\label{eq:rpf1}
\end{equation}
where $L+L'$ must be odd.  Specifically, the $R$-matrix elements relevant
for the calculation of the asymmetry are: $R^{11,0}_{00,11}$ and
$\overline{R}^{21,0}_{00,11}$ for $J=0$, and $R^{11,1}_{01,10}$,
$R^{11,1}_{01,11}$, $R^{11,1}_{21,10}$, $R^{11,1}_{21,11}$,
$\overline{R}^{21,1}_{01,10}$, and $\overline R^{21,1}_{01,11}$ for $J=1$.
Quantum Monte Carlo (QMC) techniques are employed to evaluate
these matrix elements (see below).

The asymmetry in Eq.~(\ref{eq:aaz1}) is expressed in terms of $T$-matrix elements,
which are in turn derived from $R$-matrix elements via Eq.~(\ref{eq:t21c}).  This latter
equation can be further simplified by retaining only linear terms in the PV $R$-matrix
elements, and the resulting expressions for the PC $T^{21,0}_{00,00}$,
$T^{21,1}_{01,01}$ and $T^{21,1}_{01,21}$ and PV  $T^{21,0}_{00,11}$,
$T^{21,1}_{01,10}$ and $T^{21,1}_{01,11}$ matrix elements are listed
in Appendix~\ref{app:a1}.

The QMC techniques used to evaluate the matrix element in Eq.~(\ref{eq:rpf1})
are similar to those discussed  in Ref.~\cite{Schiavilla08} for the neutron
spin rotation in $\vec{n}\, d$ scattering.  The wave functions for an assigned
spatial configuration specified by the set of Jacobi variables $({\bf x}_1,{\bf x}_2,{\bf x}_3)$
are expanded on a basis of $16 \times 6$ spin-isospin states for the four nucleons as
\begin{equation}
\psi({\bf x}_1,{\bf x}_2,{\bf x}_3) = \sum_{a=1}^{96} \psi_a({\bf x}_1,{\bf x}_2,{\bf x}_3) \mid\! a\rangle \ ,
\end{equation}
where the components $ \psi_a({\bf x}_1,{\bf x}_2,{\bf x}_3)$ are generally complex
functions, and the basis states $\mid\! a\rangle$=
$\mid\! (n\! \downarrow)_1 (p\! \downarrow)_2 (n\! \downarrow)_3 (p\! \downarrow)_4\rangle$,
$\mid\! (n\! \downarrow)_1 (n\! \downarrow)_2 (p\! \downarrow)_3 (p\! \downarrow)_4\rangle$, and
so on.  Matrix elements of the PV potential components are written
schematically as
\begin{eqnarray}
\!\!\!\langle f\!\mid O \mid \! i \rangle&=&\sum_{a,b=1}^{96} \int 
{\rm d}{\bf x}_1 \, {\rm d}{\bf x}_2 {\rm d}{\bf x}_3
\, \psi^*_{f,a}({\bf x}_1,{\bf x}_2,{\bf x}_3) \nonumber \\
\!\!\!&&\times \left[ O({\bf x}_1,{\bf x}_2,{\bf x}_3)\right]_{ab} 
\psi_{i,b}({\bf x}_1,{\bf x}_2,{\bf x}_3) \ ,
\label{eq:mci}
\end{eqnarray}
where $\left[ O({\bf x}_1,{\bf x}_2,{\bf x}_3)\right]_{ab}$ denotes the matrix representing in
configuration space any of the components in Table~\ref{tb:tab1}.  Note
that the operators ${\bf X}^{(n)}_{ij,\mp}$ occurring in $v^{\rm PV}_{ij}$
are conveniently expressed as
\begin{eqnarray}
\label{eq:xcomm}
 {\bf X}_{ij,+}^{(n)}&= &-{\rm i} \left[ 2\, f_n(r_{ij}) \,{\bm \nabla}_{{ij}} +
\hat{\bf r}_{ij}\, f^\prime_n(r_{ij}) \right] \ ,\\
  {\bf X}_{ij,-}^{(n)}&=&\hat{\bf r}_{ij} \, f^\prime_n(r_{ij}) \ , 
\end{eqnarray}
where the gradient operator ${\bm \nabla}_{ij}= ({\bm \nabla}_i-{\bm \nabla}_j)/2$
acts on the right (initial) wave function, and $f^\prime(x) ={\rm d}f(x)/{\rm
d}x$.  Gradients are discretized as
\begin{eqnarray}
\nabla_{i,\alpha} \psi ({\bf x}_1,{\bf x}_2,{\bf x}_3)\!\!&\simeq&\!\!
\big[ \psi(\dots {\bf r}_{i}+\delta\, \hat{\bf e}_\alpha \dots) \nonumber \\
\!\!&&\!\! -\psi(\dots {\bf r}_{i}-\delta\, \hat{\bf e}_\alpha \dots)\big]/(2\,\delta) \ ,
\end{eqnarray}
where $\delta$ is a small increment and $\hat{\bf e}_\alpha$ is a unit
vector in the $\alpha$-direction.  Matrix multiplications in the
spin-isospin space are performed exactly with the techniques developed
in Ref.~\cite{Schiavilla89}.  The problem is then reduced to the evaluation
of the spatial integrals, which is efficiently carried out by a combination of
MC and standard quadratures techniques.  We write
\begin{equation}
 \langle f\!\mid O \mid \! i \rangle\! = \!\int{\rm d} \hat{\bf x}_1
\,{\rm d} {\bf x}_2  {\rm d} {\bf x}_3 \, F(\hat{\bf x}_1,{\bf x}_2,{\bf x}_3) 
\! \simeq\! \frac{1}{N_c} \sum_{c=1}^{N_c} \frac{F(c)}{W(c)} \ ,
\end{equation}
where the $c$'s denote collectively (uniformly sampled)
directions $\hat{\bf x}_1$ and Jacobi coordinates $({\bf x}_2,{\bf x}_3)$,
and the probability density $W(c)=\mid \!\!\Psi({\bf x}_2,{\bf x}_3)\!\!\mid^2\!\!/(4\pi)$---$\Psi({\bf x}_2,{\bf x}_3)$
is the triton bound-state wave function normalized to one---is sampled
via the Metropolis algorithm.
For each such configuration $c$ (total number $N_c$), the function $F$ is obtained by Gaussian
integrations over the $x_1$ variable, {\it i.e.}
\begin{eqnarray}
F(c)&=& \sum_{a,b=1}^{96} \int_0^\infty {\rm d}x_1 \, x_1^2 \, 
\psi^*_{f,a}({\bf x}_1,{\bf x}_2,{\bf x}_3) \nonumber \\
&&\times \left[ O({\bf x}_1,{\bf x}_2,{\bf x}_3)\right]_{ab}  
\psi_{i,b}({\bf x}_1,{\bf x}_2,{\bf x}_3) \ .
\end{eqnarray}
Convergence in the $x$ integrations requires of the
order of 50 Gaussian points, distributed over a non-uniform grid
extending beyond 20 fm, while $N_c$ of the order of a hundred thousand
is sufficient to reduce the statistical errors in the MC integration
on the PV $T$-matrix elements at the few percent level.  In this respect,
we note that these errors are computed directly, by accumulating, in the course
of the random walk, values---and their squares---for the appropriate linear combinations of $R$-matrix
elements, as given in Eqs.~(\ref{eq:t0011}) and (\ref{eq:t0110n})--(\ref{eq:t0111n})
of Appendix~\ref{app:a1}.  Because of correlations,  the errors on the $T$-matrix
elements obtained in this way are much smaller than those that would be inferred
from the $R$-matrix elements by naive error propagation.

The present method turns out to be computationally intensive, particularly
because of the large number of wave functions (and their derivatives) that
have to be generated at each configuration $({\bf x}_1,{\bf x}_2,{\bf x}_3)$.
The computer codes have been successfully tested by carrying out
a calculation based on Gaussian wave functions for the initial
and final states, as described in the following subsection.

\subsection{Test calculation} 
\label{sec:calt}

In order to test the computer programs based on QMC techniques, we
carried out a preliminary calculation using wave functions
for which it is possible to evaluate the matrix elements of the PV potential also
analytically.  These (antisymmetric) wave functions are written as
\begin{eqnarray}
 \Psi^{JJ_z}_{\gamma, LS}&=&{1\over 4\pi} \sum_{p=1}^4 e^{-\beta \rho^2}
y_p^{L+2\, n_\beta} \nonumber \\
&\times& \biggl[ Y_L(\hat {\bf y}_p)
 \otimes \Bigl[\phi_{\gamma }(ijk) \otimes\chi_\gamma (l)\Bigr]_S \biggr]_{JJ_z} \ ,
  \label{eq:test1}
\end{eqnarray}
where $\phi_\gamma$ ($\chi$) represents a three-nucleon (single-nucleon) spin-isospin one-half
state with isospin projection --1/2 (+1/2) for $\gamma =1$ ($p$-$^3$H channel) and
+1/2 (--1/2) for $\gamma=2$ ($n$-$^3$He channel).  Thus, as in the realistic
case, the wave functions above do not have a definite total
isospin $T$ but, rather, are combinations of $T=0$ and $T=1$ states (having, of course,
$T_z=0$).  The whole radial dependence is given by the factor $y_p^{L+2\, n_\beta}
e^{-\beta \rho^2}$, where $\rho$ is the hyperradius.  The non-negative integer
$n_\beta$ and the real parameter $\beta$ can be varied so as to obtain
a family of wave functions.  For the purpose of computing matrix elements
of two-body operators, it is convenient to express the pieces in Eq.~(\ref{eq:test1})
corresponding to permutations $p \neq 1$ in terms of quantities
relative to the permutation $p=1$ or $(123,4)$.  This can be accomplished by
making use of the properties of Wigner coefficients:
\begin{widetext}
\begin{eqnarray}
  \Psi^{JJ_z}_{\gamma, LS}&=& e^{-\beta \rho^2} \sum_{\mu}
       C^{LSJ}_{\gamma\, n_\beta; \mu}\,  x_1^{n_1}  x_2^{n_2}
       x_3^{n_3} \Biggl[ \biggl[ \Bigl[ Y_{\ell_3} (\hat{\bf x}_3) \otimes
       \bigl[ \chi_1 \otimes\chi_2\bigr]_{S_2}\Bigr]_{j_3} \otimes
         \Bigl[ Y_{\ell_2}(\hat{\bf x}_2) \otimes \chi_{3}\Bigr]_{j_2} \biggr]_{J_2} \otimes
         \Bigl[ Y_{\ell_1}(\hat{\bf x}_1) \otimes \chi_{4}\Bigr]_{j_1} \Biggr]_{JJ_z}\nonumber \\
      &&\qquad \qquad \times \biggl[ \Bigl[ \bigl[\xi_1\otimes \xi_2\bigr]_{T_2}\otimes \xi_{3}\Bigr]_{T_3}
       \otimes \xi_4\biggr]_T\ , \qquad
   \mu \equiv \{ \ell_1\ell_2\ell_3 n_1 n_2 n_3 j_1 j_2 j_3 J_2 S_2 T_2  T_3 T\}\ , 
\label{eq:test2}
\end{eqnarray}
\end{widetext}
where $\chi_i$ and $\xi_i$ are the spin and isospin states of nucleon $i$,
${\bf x}_j$ are the Jacobi vectors corresponding to the permutation
$p=1$ and $n_1+n_2+n_3=L+2\,n_\beta$, and the $C$'s denote combinations
of Wigner coefficients.  It is now relatively simple to evaluate the matrix of the
PV potential $\sum_{i<j} v^{\rm PV}_{ij}=6\, v^{\rm PV}_{12}$~\cite{Schiavilla08}, by expressing
the wave functions as in Eq.~(\ref{eq:test2}).

As an example, we report here the results obtained for two $J=0$ wave functions.
In Table~\ref{tb:testwf}, we list the values of the quantum numbers $LSJ$, and
parameters $n_\beta$ and $\beta$, used in the actual calculation.  The ket $|\gamma\rangle$
with $\gamma=1$ (2) describes a ``$p$-$^3$H'' (``$n$-$^3$He'') state.
We compute the matrix elements in two ways: i) by performing the analytical
calculation via the transformation of Eq.~(\ref{eq:test2}), and ii) by
using the QMC techniques discussed earlier.
\begin{table}[bth]
\begin{tabular}{c|c|c|c|c}
\hline
\hline
State & $J^\pi$ & $LS$  & $n_\beta$ & $\beta$ \\
\hline
 $|1\rangle$ & $0^+$ & $00$ &  $0$ & 0.25 \\
 $|2\rangle$ & $0^-$  & $11$ &  $0$ & 0.25 \\
\hline
\hline
\end{tabular}
\caption{Values of the quantum numbers and parameters for some of the test wave
functions used in this work.  See text for explanation.}
\label{tb:testwf}
\end{table}

The values for the matrix elements $-\langle 1| O_{12}^{(n)} | 2\rangle$
corresponding to the $12$ components of the DDH potential (see
Table~\ref{tb:tab1}) are reported in Table~\ref{tb:test}.
\begin{table}[bth]
\begin{tabular}{c|c|c}
\hline
\hline
 $n$ & Analytical & QMC \\
\hline
           1 & $-2.987$     &  $-3.020(15) $    \\
           2 & $\m 0.000$   &  $\m 0.000\m\m$     \\
           3 & $-0.333$     &  $-0.349(4)$     \\
           4 & $-0.264$     &  $-0.281(4)$     \\
           5 & $-0.222$     &  $-0.233(2)$     \\
           6 & $\m 0.000$   &  $\m 0.000\m\m$  \\
           7 & $\m 0.000$   &  $\m 0.000\m\m$  \\
           8 & $-0.335$     &  $-0.349(6)$     \\
           9 & $-0.143$     &  $-0.147(2)$     \\
          10 & $-0.335$     &  $-0.349(6)$     \\
          11 & $-0.286$     &  $-0.294(3)$     \\
          12 & $-0.264$     &  $-0.281(4)$     \\
\hline
\hline
\end{tabular}
\caption{Results for the real part of the (adimensional) matrix element
$-\langle 1|  O_{12}^{(n)} | 2\rangle $ calculated analytically
and by using the QMC code.  For the latter calculation,
the statistical uncertainties are reported in parentheses, and
correspond to a (rather modest) set of 5k samples.
The operators $O_{12}^{(n)}$ are those of the DDH potential,
listed in Table~\ref{tb:tab1}.}
\label{tb:test}
\end{table}
There is good agreement between the results
of the two calculations.  Note that the $n=2$ contribution
associated with an isoscalar operator as well as the $n=6,7$
contributions corresponding to isotensor operators vanish.
The test wave functions consist of a superposition of $T=0$
and $T=1$ components, and therefore it is not immediately
apparent why this should be so.  The reason for this result
becomes clear only after carrying out the decomposition of
the wave functions as in Eq.~(\ref{eq:test2}).  It comes about
because of delicate cancellations among various terms.  We find
it reassuring that these same matrix elements are seen to vanish
(within machine precision) with the QMC code.  We have verified explicitly
that the close agreement between the two calculations persists
for the matrix elements involving other pairs of states, including
those having $J=1$.

\section{Further results}
\label{sec:res}

The results for the coefficients $I^\alpha_n$ in Eq.~(\ref{eq:in}),
obtained with the (zero energy) $n$-$^3$He continuum wave functions corresponding to
the AV18, AV18/UIX, N3LO, and N3LO/N2LO strong-interaction Hamiltonians,
are reported for the DDH and pionless EFT PV potentials in Tables~\ref{tb:tab8}
and~\ref{tb:tab9}, respectively.  The subscript $n$ in $I^\alpha_n$ specifies
the operators as listed in Table~\ref{tb:tab1}, and the set of cutoff parameters
entering the modified Yukawa functions are given in Table~\ref{tb:tabpv}.
\begin{table}[bth]
\begin{tabular}{c|c|c|c|c}
\hline
\hline
      & \multicolumn{4}{c} {$I_n^{\rm DDH}$} \\
\hline
\hline
$n$ &  AV18  &  AV18/UIX  &  N3LO  & N3LO/N2LO  \\
\tableline
1 &--0.186E+00 &--0.189E+00 &--0.203E+00 & --0.113E+00  \\
2 &--0.826E--02 &--0.577E--02 &--0.608E--02 & --0.622E--02 \\
3 &+0.811E--02 &+0.864E--02 &+0.333E--02 & --0.693E--02  \\
4 &--0.620E--02 &--0.794E--02 & --0.970E--02& --0.753E--02 \\
5 &--0.800E--02 &--0.976E--02 &--0.102E--01 &  --0.781E--02\\
6 &--0.359E--03 &--0.170E--03 & --0.942E--03& +0.322E--03 \\
7 &+0.631E--03 &+0.115E--02 & --0.641E--04 & +0.703E--03 \\
8 & +0.605E--02&+0.404E--02 &--0.699E--03 &  --0.794E--02\\
9 &+0.314E--02 &+0.289E--02 & --0.171E--02&  --0.577E--02\\
10 &--0.689E--02 &--0.887E--02 &--0.115E--01 &  --0.902E--02\\
11 & --0.930E--02&--0.113E--01 & --0.123E--01&--0.940E--02 \\
12 &--0.801E--02 &--0.979E--02 & --0.115E--01&  --0.606E--02\\
\hline
\hline
\end{tabular}
\caption{The coefficient $I_n^{\rm DDH}$ corresponding to the DDH potential
components $O^{(n)}$ in combination with the AV18, AV18/UIX, N3LO, N3LO/N2LO
strong interaction Hamiltonians.  The statistical Monte Carlo errors are not shown,
but are at the most 10\% for the smallest contributions, and less than 2\%
for the largest. The $I_n^{\rm DDH}$ are in units of fm$^{-1}$.}
\label{tb:tab8}
\end{table}
\begin{table}[bth]
\begin{tabular}{c|c|c}
\hline
\hline
      & \multicolumn{2}{c} {$I_n^{\rm EFT}$} \\
\hline
\hline
$n$ &  AV18/UIX  &  N3LO/N2LO  \\
\tableline
1 & --0.195E+00 & --0.119E+00 \\
4 & --0.606E+00 & --0.391E+00 \\
6 & --0.639E--02 & +0.179E--01  \\
8 & +0.608E+00 &  --0.515E--01 \\
9 &  +0.301E+00 &  +0.426E--01  \\
\hline
\hline
\end{tabular}
\caption{The coefficient $I_n^{\rm EFT}$ corresponding to the pionless EFT potential
components $O^{(n)}$ in combination with the AV18/UIX and N3LO/N2LO
strong interaction Hamiltonians.  Note that there are no potential components
with $n$=2, 3, 5, 7, 10, 11, and 12.  The statistical Monte Carlo errors are not shown,
but are typically less than 5\%. The $I_n^{\rm EFT}$ are in units of fm$^{-1}$.}
\label{tb:tab9}
\end{table}

A quick glance at Table~\ref{tb:tab8} makes it clear that i) the
contribution of the long-range component of the DDH potential
due to pion exchange is at least a factor 15 larger
than that of any of the short-range components induced by vector-meson
exchanges, and ii) among the vector-meson exchange
contributions the isoscalar ($n=2,3$ and $n=8,9$) and isovector
($n=4,5$ and $n=8$--12) ones are comparable in magnitude
and much larger than those due to isotensor $\rho$-meson exchanges
($n=6,7$).  It is also clear that the pion-exchange contribution is fairly
insensitive to the choice of input strong-interaction
Hamiltonian (with or without the inclusion of a three-nucleon
potential), used to generate the $n$-$^3$He even and odd parity
states with $J=0$ and 1.  However, the N3LO/N2LO model stands out:
the pion-range contribution is (in magnitude) substantially smaller than that
calculated for the other models.  Moreover, the isoscalar $\rho$-meson ($\omega$-meson)
contribution corresponding to $n=3$ ($n=8$) has opposite sign than obtained
for the other (AV18 and AV18/UIX) models.  

To investigate the stability of the AV18/UIX and N3LO/N2LO
results with respect to convergence in the internal part of the
wave function, we present in Table~\ref{tb:tcs2} the coefficients
$C_\alpha^i$ entering the PV observable $a_z$ in Eq.~(\ref{eq:cddh})
for two different choices of wave functions.  The results labeled
``wf2'' were listed earlier in Table~\ref{tb:tcs}, except that
those relative to the N3LO/N2LO model are based here on a smaller number
of configurations.  These results are
obtained by including in the expansion of the internal parts of the $0^{\pm}$
and $1^\pm$ wave functions the maximum number of HH functions we have
considered in the present work.  The results corresponding to the row ``wf1''
are obtained by reducing this number: in practice, for each of the classes $K_1,\dots,K_4$
we set $K_i({\rm wf1})=K_i({\rm wf2})-2$
(see discussion in Sec.~\ref{sec:whh}).  Note also that the Monte Carlo
calculation of the ``wf1'' coefficients for the AV18/UIX model
uses a factor three smaller number
of configurations, and therefore the associated statistical errors are
substantially larger.  On the other hand, the ``wf1'' and ``wf2'' N3LO/N2LO results correspond to
the same number of configurations and indeed the same random walk.
Taking into account errors, we conclude that both AV18/UIX and N3LO/N2LO
calculations have converged.  This is not the case as far as the
N3LO/N2LO singlet scattering length is concerned.

\begin{widetext}
\begin{center}
\begin{table}[bth]
\begin{tabular}{l|c||c|c|c||c|c}
\hline
\hline
  &  $C_\pi^1$  &  $C_\rho^0$  &  $C_\rho^1$  & $C_\rho^2$ & $C_\omega^0$  &$C_\omega^1$  \\
\tableline
 AV18/UIX-wf1  & --0.2077(281)   & --0.0433(116)  &+0.0242(29)    & --0.0011(2)
 &--0.0232(77)   &  +0.0490(30)    \\ 
 AV18/UIX-wf2 &--0.1853(150) &--0.0380(70)    & +0.0230(18)   &  --0.0011(1)
 &--0.0231(56)    &   +0.0500(20) \\ 
\hline
N3LO/N2LO-wf1  & --0.1118(29)    & +0.0369(25)   & +0.0200(8)   & --0.0009(1) & +0.0390(23)  & +0.0402(12)     \\ 
N3LO/N2LO-wf2  & --0.1050(35)   &  +0.0445(33)  & +0.0189(9)  & --0.0008(1) & +0.0454(31)  &  +0.0417(12)   \\ 
\hline
\hline
\end{tabular}
\caption{The coefficients $C_\alpha^i$ entering the PV observable $a_z$, corresponding to the
AV18/UIX and N3LO/N2LO strong-interaction Hamiltonians for two sets of wave
functions (see text for details).  The statistical errors due to the Monte Carlo
integrations are indicated in parentheses.
}
\label{tb:tcs2}
\end{table}
\end{center}
\end{widetext}

Therefore, the differences found between the N3LO/N2LO
and the other models are presumably due to the fact that the HH expansion
for the N3LO/N2LO wave functions (specifically the $0^+$ wave function)
has not fully converged.  Consequently, in the following we restrict our discussion
to the results obtained with the AV18, N3LO, and
AV18/UIX models.  In reference to the pion contribution, the calculated
$C^1_\pi$ is rather insensitive to the choice of strong Hamiltonian.
However, there is still a considerable model dependence in the results obtained
for the individual contributions due to vector-meson exchanges.
This model dependence, in turn, impacts very significantly predictions for the PV
asymmetry $a_z$, as it can be surmised from Table~\ref{tb:tab10}.
Of course, this is so under the assumption that the values for the
strong- and weak-interaction coupling constants characterizing the
DDH potential are those listed in Table~\ref{tb:tabpv}.  For example,
the combination of coupling constants corresponding to pion-exchange
($n=1$) and isoscalar $\rho$-meson exchange ($n=2$ and 3) are,
respectively, $c_1^{\rm DDH}=(4.48\times 10^{-7})$ fm, $c_2^{\rm DDH}=(11.2\times10^{-7})$ fm
and $c_3^{\rm DDH}=(79.5\times10^{-7})$ fm---note that $c_3^{\rm DDH} =(1+\kappa_\rho) c_2^{\rm DDH}$
and $\kappa_\rho=6.1 $ is the value adopted here for the tensor coupling of the $\rho$-meson to the
nucleon~\cite{Machleidt01}.  Consequently, the contribution $c_3^{\rm DDH} \times I_3^{\rm DDH}$ is
comparable in magnitude and opposite in sign to the pion-exchange contribution
$c_1^{\rm DDH} \times I_1^{\rm DDH}$.  In this respect, we note that
the asymmetry $a_z$ changes roughly from $-27\times 10^{-8}$ to $ +13  \times 10^{-8}$
as the six PV weak coupling constants entering the DDH model are varied over
their respective allowed ranges determined in Ref.~\cite{Desplanques80}.  Thus,
$a_z$ could potentially be large enough to make its measurement (relatively) easy.
\begin{widetext}
\begin{center}
\begin{table}[bth]
\begin{tabular}{c|c|c|c|c}
\hline
\hline
      & \multicolumn{4}{c} {$10^{8} \times a^{\rm DDH}_z$} \\
\hline
\hline
$n$ &  AV18  &  AV18/UIX  &  N3LO  & N3LO/N2LO  \\
\tableline
 1 & --8.33$\pm$0.35  & --8.45$\pm$0.69    & --9.07$\pm$0.40 & --5.06$\pm$0.34 \\
 2 &  --9.26$\pm$0.35  &  --9.09$\pm$0.70  & --9.75$\pm$0.40  &--5.76$\pm$0.38 \\
 3 & --2.80$\pm$0.68  &   --2.22$\pm$1.34  & --7.10$\pm$0.89  & --11.3$\pm$0.98 \\
 4 &  --2.86$\pm$0.68  & --2.30$\pm$1.34  &   --7.20$\pm$0.89 & --11.3$\pm$0.98\\
 5 & --3.40$\pm$0.68  &  --2.95$\pm$1.34  &   --7.88$\pm$0.89& --11.9$\pm$0.98\\
 6 & --3.41$\pm$0.68  &   --2.95$\pm$1.34 &   --7.90$\pm$0.89 &--11.9$\pm$0.98\\
 7 &  --3.32$\pm$0.68 &   --2.80$\pm$1.34 &   --7.91$\pm$0.89 &--11.8$\pm$0.98\\
 8 &  --3.97$\pm$0.69  &   --3.23$\pm$1.35&  --7.83$\pm$0.90& --10.9$\pm$0.99\\
 9 & --4.31$\pm$0.69  &   --3.55$\pm$1.35  &  --7.65$\pm$0.90 &--10.3$\pm$0.99 \\
10&   --4.09$\pm$0.69 &  --3.26$\pm$1.35&  --7.28$\pm$0.90& --10.0$\pm$0.99\\
11&   --3.79$\pm$0.69 &   --2.89$\pm$1.35  & --6.88$\pm$0.90&--9.70$\pm$0.99\\
 12&  --3.45$\pm$0.69  &  --2.48$\pm$1.35  & --6.40$\pm$0.90& --9.44$\pm$0.99\\
\hline
variation&from --27.1 to +13.3  &from --27.6 to +13.8 & from --26.0 to +3.68 & from --29.7 to +6.66 \\
\hline
\hline
\end{tabular}
\caption{Cumulative contributions to $a_z$ and associated errors (rows 1--12), obtained for the DDH
PV potential with values for the coupling constants as listed in Table~\protect\ref{tb:tabpv}.
The four columns correspond to the different combinations of strong-interaction Hamiltonians
adopted in the calculations.  The last row shows the minimum and maximum (central) values that
$a_z$ can attain, as the PV couplings are varied over the allowed ranges in the original
DDH formulation~\protect\cite{Desplanques80}.  }
\label{tb:tab10}
\end{table}
\end{center}
\end{widetext}

The coefficients $I^{\rm EFT}_n$ for the operators
entering the pionless EFT PV potential, that is $n$=1, 4, 6, 8,
and 9, are reported in Table~\ref{tb:tab9}.  The coefficients
$I^{\rm EFT}_n$ for $n$=1, 4, 8, and 9, corresponding to isoscalar
and isovector structures, are all of the same order
of magnitude, while that for $n$=6 with isotensor character is
much smaller.  Note that the radial functions are taken to be
the same for all $n$, $f^{\rm EFT}_n(r)=f_\mu(r)$.
Of course, the $I^{\rm EFT}_n$'s will depend significantly on the value of the mass $\mu$---either
$\mu=m_\pi$, as appropriate in the present pionless EFT formulation,
or $\mu=1$ GeV, the scale of chiral symmetry breaking,
as appropriate in the formulation in which pion degrees of freedom
are explicitly retained.  Indeed, in this latter formulation
the leading order component of $v^{\rm PV}$ has the same form as
the pion-exchange term in DDH.

Finally, rough estimates have been made for the range of values
allowed for the low-energy constants $C_1$, $C_2+C_4$, $C_5$,
$\tilde C_1$, and $C_6$ in Ref.~\cite{Zhu05}.  However, at the
present time a systematic program for their determination is yet
to be carried out.  In view of this, we refrain here from making EFT-based
predictions for the longitudinal asymmetry.

\section*{Acknowledgments}
The authors would like to thank J.D.\ Bowman, C.B.\ Crawford, and M.T.\ Gericke
for their continued interest in the present work and for correspondence in reference
to various aspects of the calculations.

One of the authors (R.S.) would also like to thank the Physics Department of
the University of Pisa, the INFN Pisa branch, and especially the Pisa group
for the support and warm hospitality extended to him on several occasions.
The work of R.S.\ is supported by the U.S.~Department of Energy,
Office of Nuclear Physics, under contract DE-AC05-06OR23177.
The calculations were made possible by grants of computing
time from the National Energy Research Supercomputer Center.
\appendix
\section{From $R$- to $T$-matrices}
\label{app:a1}

Consider the case with $J=0$ first.  For the parity-conserving (PC)
$T$-matrix we have:
\begin{eqnarray}
  T^{21,0}_{00,00}&=&{1\over \sqrt{q_1}}\Big[ \overline R^{21,0}_{00,00}
 (I-i\,R^{11,0})^{-1}_{00,00} \nonumber \\
&&+   \overline R^{21,0}_{00,11}
 (I-i\, R^{11,0})^{-1}_{11,00}\Big]\ , \label{eq:t0000}
\end{eqnarray}
where $I-i\, R^{11}$ is a $2\times2$ matrix with very small off-diagonal elements, {\it i.e.}
\begin{eqnarray}
&& I-i\, R^{11,0}= \left( \begin{array}{cc}
            a & \epsilon \\
            \epsilon & b \\
             \end{array}   \right)\ , \,\,\,\,  a=1-i\, R^{11,0}_{00,00}\ , \nonumber \\
&&\,\,\,\, \epsilon=-i\,
           R^{11,0}_{00,11}\ ,\,\,\,\,  b=1-i\, R^{11,0}_{11,11}\ , \label{eq:r11}
\end{eqnarray}
with $|a|, |b| \gg |\epsilon|$.  To first order in $\epsilon$, we approximate
\begin{equation}
  (I-i\, R^{11,0})^{-1}= \left( \begin{array}{cc}
            1/a & -\epsilon/ab \\
            -\epsilon/ab & 1/b \\
             \end{array}   \right)\ ,\label{eq:r11i}
\end{equation}
and hence
\begin{equation}
  T^{21,0}_{00,00}={1\over \sqrt{q_1}} { \overline{R}^{21,0}_{00,00}\over a} \ .\label{eq:t0000b}
\end{equation}
Similarly, for the parity-violating (PV) $T$-matrix element we find:
\begin{widetext}
\begin{equation}
  T^{21,0}_{00,11}={1\over \sqrt{q_1}}\Bigl[ \overline R^{21,0}_{00,00}
  (I-i\, R^{11,0})^{-1}_{00,11}+   \overline R^{21,0}_{00,11}
  (I-i\, R^{11,0})^{-1}_{11,11}\Bigr] 
   ={1\over \sqrt{q_1}}\left[
  {i\,  \overline R^{21,0}_{00,00}\, R^{11,0}_{00,11}\over a\, b} +   {\overline R^{21,0}_{00,11}
  \over b} \right]\ .\label{eq:t0011}
\end{equation}
\end{widetext}
The case $J=1$ is somewhat more involved since the matrices are now $4\times4$.
The matrix $(I-i\, R^{11,1})^{-1}$ is written as
    \begin{equation}
      I-i\, R^{11,1}=\left( \begin{array}{cc}
            A & \epsilon \\
            \epsilon^T & B \\
             \end{array}   \right)\ .\label{eq:r11J}
    \end{equation}
 where $A$, $\epsilon$, and $B$ are $2\times2$ matrices,
    \begin{eqnarray}
      A&=&\left( \begin{array}{cc}
            1-i\, R^{11,1}_{01,01} & -i\, R^{11,1}_{01,21} \\
            -i\, R^{11,1}_{21,01}& 1-i\, R^{11,1}_{21,21} \\
             \end{array}   \right)\ ,\label{eq:mata}\\
      B&=&\left( \begin{array}{cc}
            1-i\, R^{11,1}_{10,10} & -i\, R^{11,1}_{10,11} \\
            -i\, R^{11,1}_{11,10}& 1-i\, R^{11,1}_{11,11} \\
             \end{array}   \right)\ ,\label{eq:matb}\\
      \epsilon&=&\left( \begin{array}{cc}
            -i\, R^{11,1}_{01,10} & -i\, R^{11,1}_{01,11} \\
            -i\, R^{11,1}_{21,10}& -i\, R^{11,1}_{21,11} \\
             \end{array}   \right)\ .\label{eq:mate}
    \end{eqnarray}
Note that $A$ and $B$, as well as their inverse $A^{-1}$ and $B^{-1}$,
are symmetric.  To first order in $\epsilon$, it follows that
    \begin{equation}
      (I-i\, R^{11,1})^{-1}=\left( \begin{array}{cc}
            A^{-1} &C \\
            C^T & B^{-1} \\
             \end{array}   \right)\ ,\label{eq:r11J3}
    \end{equation}
where the  $2\times2$ matrix $C$ and its transpose are defined as
\begin{equation}
C=-A^{-1}\epsilon B^{-1}\ , \qquad C^T=-B^{-1} \epsilon^T A^{-1} \ .
\end{equation}
This shows that $(I-i\, R^{11,1})^{-1}$ is also symmetric in this approximation.
The PC $T^{21,1}_{01,01}$ and
 $T^{21,1}_{01,21}$ and PV  $T^{21,1}_{01,10}$ and $T^{21,1}_{01,11}$ matrix
 elements entering Eq.~(\ref{eq:t21c}) are then given by
  \begin{eqnarray}
    T^{21,1}_{01,01}\!\!&=&\!\!{1\over \sqrt{q_1}} \Bigl[ \overline R^{21,1}_{01,01}
    (A^{-1})_{01,01}\nonumber \\
&&\qquad +\,
     \overline R^{21,1}_{01,21} 
(A^{-1})_{21,01}\Bigr]  \ ,\label{eq:t0001n} \\ 
    T^{21,1}_{01,21}\!\!&=&\!\!{1\over \sqrt{q_1}} \Bigl[ \overline R^{21,1}_{01,01}
   (A^{-1})_{01,21} \nonumber \\
&&\qquad+\,
     \overline R^{21,1}_{01,21} (A^{-1})_{21,21}\Bigr]  \ ,\label{eq:t0021n} \\
    T^{21,1}_{01,10}\!\!&=&\!\!{1\over \sqrt{q_1}}\Bigl[ \overline R^{21,1}_{01,01} C_{01,10}
    +\, \overline R^{21,1}_{01,21} C_{21,10} 
 + \,  \overline R^{21,1}_{01,10} \nonumber \\
&&\times (B^{-1})_{10,10}
         + \,  \overline R^{21,1}_{01,11}(B^{-1})_{11,10}
      \Bigr]\ ,\label{eq:t0110n} \\
    T^{21,1}_{01,11}\!\!&=&\!\!{1\over \sqrt{q_1}}\Bigl[ \overline R^{21,1}_{01,01} C_{01,11}
    + \overline R^{21,1}_{01,21} C_{21,11} +
    \overline R^{21,1}_{01,10}\nonumber \\
&&\times (B^{-1})_{10,11}
         +   \overline R^{21,1}_{01,11} (B^{-1})_{11,11}
      \Bigr]\ .\label{eq:t0111n}
  \end{eqnarray}
%
%
\section{Numerical values for $R$- and $T$-matrix elements}
\label{app:a2}

The set of Tables~\ref{tb:tab4}--\ref{tb:tab7} are all relative to the
AV18/UIX+DDH model, and present results for the $R$-matrix elements
involving PV transitions between states with $J=0$ and $J=1$, the corresponding
$T$-matrix elements which follow from them and the parity-conserving (PC) $R$-matrix
elements via Eqs.~(\ref{eq:t0011}) and~(\ref{eq:t0110n})--(\ref{eq:t0111n}), and lastly
the coefficients $d^{(n)}_i$,
\begin{table}[bth]
\begin{tabular}{c|c|c}
\hline
\hline
$n$ & $R^{11,0}_{00,11}$  & $\overline{R}^{21,0}_{00,11}$   \\
\tableline
1 &  +0.198E+01  & +0.278E+01  \\
2 & +0.126E+00   & +0.305E+00  \\
3 & --0.149E+00 &--0.373E+00   \\
4 &  +0.533E--01  & +0.530E--01 \\
5 &  +0.632E--01  & +0.691E--01   \\
6 &  +0.156E--02  & +0.154E--02 \\
7 &   +0.129E-02 & +0.163E--02 \\
8 &   --0.211E+00 & --0.523E+00 \\
9 &   --0.797E--01& --0.203E+00   \\
10 &  +0.589E--01  &+0.588E--01    \\
11 &  +0.720E--01  & +0.784E--01  \\
12 &  +0.154E--01 & +0.134E--01  \\
\hline
\hline
\end{tabular}
\caption{The parity-violating $R$-matrix elements for $J=0$ corresponding to the DDH potential
components $O^{(n)}$ in combination with the AV18/UIX strong interaction potentials at
vanishing $n$-$^3$He energy.  The statistical Monte Carlo errors are not shown,
but are typically $\sim$ 1--2 \% for the largest values, and less than 10\% for the smallest.
The $R$-matrix element without (with) overline is in units of fm$^{-1}$ (fm$^{-1/2}$), see
text for explanation. }
\label{tb:tab4}
\end{table}
\begin{eqnarray}
d^{(n)}_1 &=&\overline T^{21,1}_{01,10}(n)\,  \overline T^{21,0\, *}_{00,00} \ , \qquad
d^{(n)}_2\,=\,\overline T^{21, 0 }_{00,11}(n)\, \overline T^{21,1\, *}_{01,01} \ , \nonumber \\
d^{(n)}_3&=&\overline T^{21,0 }_{00,11}(n)\,  \overline T^{21,1\, *}_{01,21}\ ,\qquad
d^{(n)}_4\,=\, \overline T^{21,1}_{01,11}(n)\,  \overline T^{21,1\, *}_{01,01}\ , \nonumber \\
d^{(n)}_5&=&\overline T^{21,1 }_{01,11}(n)\, \overline T^{21,1\, * }_{01,21} \ ,
\end{eqnarray}
where the $\overline{T}$-matrix elements are defined as in Eq.~(\ref{eq:tbar}), and the label
$(n)$ on those involving PV transitions refers to the operator component $O^{(n)}$ in Table~\ref{tb:tab1}.
The $I_n$'s discussed earlier follow from
\begin{eqnarray}
I_n&=& -{4\over \Sigma}\, {\rm Re}
\Big[ \sqrt{3}\, d_1^{(n)} - d_2^{(n)} +\sqrt{2}\, d_3^{(n)}
\nonumber \\
&&+\sqrt{6}\,d_4^{(n)}
+\sqrt{3}\, d_5^{(n)}\Big]\  ,
\end{eqnarray}
where $\Sigma$ has been defined in Eq.~(\ref{eq:ssss}).  A few words on units:
since the operators $O^{(n)}$ do not include
the $c_n$'s, {\it i.e.} the combinations  of nucleon mass and
strong- and weak-interaction coupling constants, the resulting
$R$-matrix ($T$-matrix) elements involving PV transitions
are in units of fm$^{-1}$ (adimensional)---they would otherwise be adimensional
(in units of fm).  Further, because of the definition in Eq.~(\ref{eq:rbar}), the
$\overline{R}$-matrix elements have dimensions of fm$^{-1/2}$.  Note, however,
that the $\overline{T}$- and $T$-matrix elements only differ by a phase factor, and
hence the former are also adimensional.
\begin{widetext}
\begin{center}
\begin{table}[bth]
\begin{tabular}{c|c|c|c|c|c|c}
\hline
\hline
$n$ & $R^{11,1}_{01,10}$  & $R^{11,1}_{01,11}$ &
  $R^{11,1}_{21,10}$ & $R^{11,1}_{21,11}$ & $\overline{R}^{21,1}_{01,10}$ &  $\overline R^{21,1}_{01,11}$ \\
\tableline
1 &  --0.160E--01   & --0.930E--01  & +0.199E--02    & +0.365E--02    & --0.535E--01 & --0.106E--01 \\
2 & +0.614E--03     &  +0.131E--02  & --0.216E--04   &--0.627E--04   &--0.339E--02  & +0.854E--02 \\
3 & --0.837E--03   & --0.198E--02  &  --0.204E--04 & +0.941E--04  & +0.528E--02 & +0.604E--03 \\
4 &  --0.188E--03   & +0.782E--03  & --0.643E--04   & --0.958E--05  & --0.179E--02  &+0.161E--02  \\
5 &   --0.317E--03  & +0.918E--03  &--0.853E--04   & --0.133E--04  & --0.232E--02 &  +0.186E--02\\
6 &    +0.116E--03  &+0.159E--02   & --0.404E--04   & --0.777E--04  & --0.257E--03 & --0.427E--02 \\
7 &    --0.186E--04   &+0.191E--02   & --0.713E--04  &--0.870E--04   &+0.617E--04   &--0.518E--02  \\
8 &   --0.769E--03   &--0.181E--02   & --0.217E--04  & +0.860E--04   &  +0.506E--02& +0.812E--03 \\
9 &   --0.364E--03    &--0.852E--03    &--0.274E--04   & +0.382E--04   &+0.260E--02   &+0.448E--02  \\
10 &  --0.211E--03    & +0.867E--03  & --0.716E--04  &--0.107E--04   &--0.201E--02  &+0.179E--02  \\
11 &   --0.367E--03   & +0.105E--02  &--0.985E--04   &  --0.151E--04 &--0.270E--02  &  +0.214E--02 \\
12 &   --0.543E--03  &--0.144E--02   &  --0.699E--05 &  +0.636E--04  & --0.258E--02  &  --0.102E--03\\
\hline
\hline
\end{tabular}
\caption{Same as in Table~\ref{tb:tab4}, but for $J=1$.}
\label{tb:tab5}
\end{table}
\end{center}
\begin{center}
\begin{table}[bth]
\begin{tabular}{c|c|c||c|c||c|c}
\hline
\hline
      & \multicolumn{2}{c} {$T^{21,0}_{00,11}$} & \multicolumn{2}{c}{$T^{21,1}_{01,10}$} &
           \multicolumn{2}{c} {$T^{21,1}_{01,11}$} \\
\hline
\hline
$n$ &  Re  &  Im  &  Re  & Im & Re  & Im  \\
\tableline
1 & --0.104E+01  & --0.302E+01  & --0.133E+00 & --0.134E--02  &   --0.316E--01 & --0.168E--01   \\
2 & +0.219E+00   &  --0.996E--01 &--0.830E--02  &  +0.143E--03  & +0.210E--01    & +0.123E--02  \\
3 & --0.289E+00 & +0.108E+00  & +0.129E--01 & --0.269E--03  & +0.136E--02    &  --0.293E--03 \\
4 & --0.767E--01 &  --0.971E--01 & --0.442E--02 &  +0.335E--05 &  +0.401E--02  &  +0.330E--03 \\
5 &  --0.771E--01 & --0.111E+00  &--0.573E--02  &--0.682E--05   &  +0.463E--02  &  +0.385E--03  \\
6 &  --0.226E--02 & --0.285E--02  & --0.625E--03 &  +0.453E--04 &  --0.104E--01  &  --0.224E--03 \\
7 & --0.110E--02  & --0.210E--02  & +0.150E--03  & +0.194E--04  &  --0.126E--01  &  --0.277E--03 \\
8 & --0.393E+00 & +0.159E+00  & +0.124E--01  & --0.253E--03   &  +0.189E--02  &  --0.239E--03 \\
9 & --0.161E+00 & +0.559E--01  & +0.637E--02 & --0.144E--03  & +0.110E--01    & +0.364E--03  \\
10 &--0.843E--01  &  --0.107E+00 &  --0.495E--02 & +0.375E--05  & +0.445E--02     & +0.367E--03  \\
11 & --0.887E--01 & --0.126E+00  & --0.667E--02  & --0.757E--05   & +0.533E--02    &  +0.443E--03  \\
12 & --0.265E--01 & --0.295E--01  &  --0.640E--02 &--0.285E--04   & --0.340E--03    &  --0.245E--03 \\
\hline
\hline
\end{tabular}
\caption{The parity-violating $T$-matrix elements corresponding to the DDH potential
components $O^{(n)}$ in combination with the AV18/UIX strong interaction potentials at
vanishing $n$-$^3$He energy.  The statistical Monte Carlo errors are not shown,
but are typically less than 10\%. The $T$-matrix elements are adimensional,
see text for explanation.}
\label{tb:tab6}
\end{table}
\end{center}
\begin{center}
\begin{table}[bth]
\begin{tabular}{c|c|c|c|c|c||c}
\hline
\hline
$n$ &  Re$\, d^{(n)}_1$  &  Re$\, d^{(n)}_2$  &  Re$\, d^{(n)}_3$  & Re$\, d^{(n)}_4$ & Re$\, d^{(n)}_5$  & $I^{\rm DDH}_n$  \\
\tableline
1  &+0.617E+00  &    +0.349E--01 &     +0.107E--01 &    --0.414E--02   &   +0.616E--04  &    --0.189E+00 \\
 2 &+0.384E--01    &  +0.436E--01   &   +0.333E--03   &   +0.345E--02     &--0.597E--05     &--0.577E--02 \\
 3&--0.598E-- 01    & --0.559E--01   &  --0.356E--03   &   +0.249E--03    &  +0.923E--06    &  +0.864E--02 \\
 4&+0.205E--01    & --0.614E--02    &  +0.347E--03    &  +0.651E--03     &--0.147E--05     &--0.794E--02 \\
5&+0.266E--01    & --0.527E--02     & +0.395E--03     & +0.751E--03     &--0.171E--05    & --0.976E--02 \\
6&+0.287E--02     &--0.183E--03      &+0.102E--04     &--0.173E--02      &+0.159E--05     &--0.170E--03 \\
7&--0.709E--03    & --0.378E--04     & +0.748E--05    & --0.210E--02      &+0.195E--05      &+0.115E--02 \\
8&--0.573E--01    & --0.769E--01     &--0.528E--03     & +0.333E--03     & +0.694E--06      &+0.404E--02 \\
9&--0.294E--01    & --0.309E--01     &--0.184E--03     & +0.181E--02     &--0.213E--05      &+0.289E--02 \\
10&+0.230E--01     &--0.672E--02      &+0.383E--03      &+0.722E--03     &--0.163E--05     &--0.887E--02 \\
11&+0.310E--01    & --0.612E--02      &+0.451E--03     & +0.864E--03     &--0.197E--05     &--0.113E--01 \\
12 &+0.297E--01     &--0.241E--02      & +0.106E--03     &--0.401E--04     & +0.888E--06    & --0.979E--02 \\
\hline
\hline
\end{tabular}
\caption{The real parts of the coefficients $d_i$ ($i=1, \dots, 5$),
and the coefficients $I_n^{\rm DDH}$, corresponding to the DDH potential
components $O^{(n)}$ in combination with the AV18/UIX strong interaction potentials.
The statistical Monte Carlo errors are not shown, but are typically less than 10\%. The 
$d_i$ are adimensional, while $I_n^{\rm DDH}$ is in units of fm$^{-1}$.}
\label{tb:tab7}
\end{table}
\end{center}
\end{widetext}
The $R$-matrix elements in $J=1$ states (Table~\ref{tb:tab5}) are typically two orders of magnitude smaller
than those in $J=0$ states (Table~\ref{tb:tab4}).  Among the former, those with orbital angular momentum
$L=2$ in channel $p$-$^3$H ($\gamma=1$) are much suppressed at the low energies
of interest in the present work. Inspection of Table~\ref{tb:tab4} also shows that the
(isovector) pion-exchange interaction ($n=1$) is dominant, which suggests that
the $J^\pi=0^+$ and $0^-$ states in both $n$-$^3$He and $p$-$^3$H are not
purely isoscalar, but rather have significant admixtures of isospin components $T>0$.

In order to compute the $d_i$'s in Table~\ref{tb:tab7}, one needs, in addition to
the $T$-matrix elements listed in Table~\ref{tb:tab6}, also the $T$-matrix elements
associated with PC transitions.  These have been calculated to be (at zero $n$-$^3$He
energy): $T_{00,00}^{21,0}=(-1.356+i\, 4.482)$ fm, $T_{01,01}^{21,1}=(0.1679-i\, 0.6937)$ fm,
and $T_{01,21}^{21,1}=(0.003497-i\, 0.0003535)$ fm.
We conclude by noting that the $d_1^{(n)}$ and $d_2^{(n)}$ combinations
give the leading contributions to $I_n$ and that, in the case of pion exchange,
$d^{(1)}_1$ is in fact dominant.  This fits in well with the expectation that the
$^1$S$_0 \rightarrow \,^3$P$_0$ transition entering
$\overline{T}^{21,0}_{00,11}$ in $d_2^{(1)}$ is predominantly isoscalar,
while $d_1^{(1)}$ involves the transition $^3$S$_1 \rightarrow\,^1$P$_1$
in $\overline{T}^{21,1}_{01,10}$, which
presumably has both isoscalar and isovector character.  Indeed,
the contributions of isoscalar $\rho$- and $\omega$-exchange interactions
are comparable in $d^{(2)}_1$, $d^{(3)}_2$ and $d^{(8)}_1$, $d^{(9)}_2$,
respectively.


\begin{thebibliography}{100}
%
\bibitem{Haxton08}
W.C.\ Haxton,
arXiv:0802.2984 [nucl-th].
%
\bibitem{Wiringa02}
R.B.\ Wiringa and S.C.\ Pieper,
Phys.\ Rev.\ Lett.\ {\bf 89}, 182501 (2002).
%
\bibitem{Otsuka05}
T.\ Otsuka, T.\ Suzuki, R.\ Fujimoto, H.\ Grawe, and Y.\ Akaishi,
Phys.\ Rev.\ Lett. {\bf 95}, 232502 (2005);
T.\ Otsuka, T.\ Matsuo, and D.\ Abe,
Phys.\ Rev.\ Lett.\ {\bf 97}, 162501 (2006).
%
\bibitem{Schiffer04}
J.P.\ Schiffer {\it et al.},
Phys.\ Rev.\ Lett.\ {\bf 92}, 162501 (2004).
%
\bibitem{Schiavilla07}
R.\ Schiavilla, R.B.\ Wiringa, S.C.\ Pieper, and J.\ Carlson,
Phys.\ Rev.\ Lett.\ {\bf  98}, 132501 (2007).
%
\bibitem{Subedi08}
R.\ Subedi {\it et al.},
Science {\bf 320}, 1476 (2008);
R.\ Shneor {\it et al.},
Phys.\ Rev.\ Lett. {\bf 99}, 072501 (2007).
%
\bibitem{Carlson02}
J.\ Carlson, R.\ Schiavilla, V.R.\ Brown, and B.F.\ Gibson,
Phys.\ Rev.\ C {\bf 65}, 035502 (2002).
%
\bibitem{Schiavilla04}
R.\ Schiavilla, J.\ Carlson, and M.\ Paris,
Phys.\ Rev.\ C {\bf 70}, 044007 (2004).
%
\bibitem{Schiavilla08}
R.\ Schiavilla, M.\ Viviani, L.\ Girlanda, A.\ Kievsky, and L.E.\ Marcucci,
Phys.\ Rev.\ C {\bf 78}, 014002 (2008).
%
\bibitem{Arriaga10}
A.\ Arriaga, J.\ Carlson, K.M.\ Nollett, R.\ Schiavilla, M.\ Viviani, and R.B.\ Wiringa,
in preparation.
%
\bibitem{Balzer80}
R.\ Balzer {\it et al.},
Phys.\ Rev.\ Lett.\ {\bf 44}, 699 (1980).
%
\bibitem{Yuan86}
V.\ Yuan {\it et al.},
Phys.\ Rev.\ Lett.\ {\bf 57}, 1680 (1986).
%
\bibitem{Eversheim91}
P.D.\ Eversheim {\it et al.},
Phys.\ Lett.\ {\bf B256}, 11 (1991).
%
\bibitem{Berdoz03}
A.R.\ Berdoz {\it et al.},
Phys.\ Rev.\ C {\bf 68}, 034004 (2003).
%
\bibitem{Lang85}
J.\ Lang {\it et al.},
Phys.\ Rev.\ Lett.\ {\bf 54}, 170 (1985).
%
\bibitem{Cavaignac77}
J.F.\ Cavaignac, B.\ Vignon, and R.\ Wilson,
Phys.\ Lett.\ {\bf B67}, 148 (1977).
%
\bibitem{Gericke09}
M.T.\ Gericke {\it et al.},
Nucl.\ Inst.\ Meth.\ {\bf A611}, 239 (2009).
%
\bibitem{Knyazkov84}
V.A.\ Knyaz'kov {\it et al.},
Nucl.\ Phys.\ {\bf A417}, 209 (1994).
%
\bibitem{Snow09}
W.M.\ Snow,
Nucl.\ Inst.\ Meth.\ {\bf A611}, 248 (2009).
%
\bibitem{Bass09}
C.D.\ Bass {\it et al.},
Nucl.\ Inst.\ Meth.\ {\bf A612}, 69 (2009).
%
\bibitem{Markoff07}
D.M.\ Markoff,
private communication.
%
\bibitem{Bowman07}
J.D.\ Bowman (PI), C.B.\ Crawford (PI), M.T.\ Gericke (PI) {\it et al.},
{\it A Measurement of the Parity Violating Proton Asymmetry in the Capture of Polarized
Cold Neutrons on 3He}, proposal submitted to the SNS FNPB PRAC~T, 2007-11-25.
%
\bibitem{Kievsky08} A.\ Kievsky {\it et al.}, J. Phys. G:
   Nucl. Part. Phys. {\bf 35}, 063101 (2008).
%
\bibitem{Viviani10}
M.\ Viviani {\it et al.}, in preparation.
%
\bibitem{Wiringa95}
R.B.\ Wiringa, V.G.J.\ Stoks, and R.\ Schiavilla,
Phys.\ Rev.\ C {\bf 51}, 38 (1995).
%
\bibitem{Entem03}
D.R.\ Entem and R.\ Machleidt,
Phys.\ Rev.\ C {\bf 68}, 041001 (2003).
%
\bibitem{Pudliner97}
B.S.\ Pudliner, V.R.\ Pandharipande, J.\ Carlson, S.C.\ Pieper, and R.B.\ Wiringa,
Phys.\ Rev.\ C {\bf 56}, 1720 (1997).
%
\bibitem{Navratil07} 
P. Navr{\'a}til,  Few-Body Syst. {\bf 41}, 117 (2007);
D.\ Gazit, S.\ Quaglioni, and P.\ Navratil,
Phys.\ Rev.\ Lett.\ {\bf 103}, 102502 (2009).
%
\bibitem{Desplanques80}
B.\ Desplanques, J.F.\ Donoghue, B.R.\ Holstein,
Ann.\ Phys.\ (N.Y.) {\bf 124}, 449 (1980).
%
\bibitem{Zhu05}
S.-L.\ Zhu, C.M.\ Maekawa, B.R.\ Holstein, M.J.\ Ramsey-Musolf, and U.\ van Kolck,
Nucl.\ Phys.\ {\bf A748}, 435 (2005).
%
\bibitem{Girlanda08}
L.\ Girlanda,
Phys.\ Rev.\ C {\bf 77}, 067001 (2008).
%
\bibitem{Jurney82}
E.T.\ Jurney, P.J.\ Bendt, and J.C.\ Browne,
Phys.\ Rev.\ C {\bf 25}, 2810 (1982);
M.W.\ Konijnenberg {\it et al.},
Phys.\ Lett.\ {\bf B205}, 215 (1988).
%
\bibitem{Wolfs89}
F.L.H.\ Wolfs, S.J.\ Freedman, J.E.\ Nelson, M.S.\ Dewey, and G.L.\ Greene,
Phys.\ Rev.\ Lett.\ {\bf 63}, 2721 (1989);
R.\ Wervelman, K.\ Abrahams, H.\ Postma, J.G.L.\ Booten, and A.G.M.\ Van Hees,
Nucl.\ Phys.\ {\bf A526}, 265 (1991).
%
\bibitem{Girlanda10}
L.\ Girlanda, S.\ Pastore, R.\ Schiavilla, and M.\ Viviani,
EPJ Web of Conferences {\bf 3}, 01004 (2010);
L.\ Girlanda, A.\ Kievsky, L.E.\ Marcucci, S.\ Pastore, R.\ Schiavilla, and M.\ Viviani,
in preparation.
%
\bibitem{Pastore09}
S.\ Pastore, L.\ Girlanda, R.\ Schiavilla, M.\ Viviani, and R.B.\ Wiringa,
Phys.\ Rev.\ C {\bf 80}, 034004 (2009).
%
\bibitem{Schiavilla87}
R.\ Schiavilla and V.R.\ Pandharipande,
Phys.\ Rev.\ C {\bf 36}, 2221 (1987).
%
\bibitem{Note1}
In principle, there is one additional parity-violating coupling constant, $h_\rho^{1\, '}$,
multiplying an isovector operator of $\rho$-meson range.  However, theoretical
estimates~\protect\cite{Holstein81} indicate that its value is much smaller than the values
obtained for the other coupling constants in the original DDH reference~\protect\cite{Desplanques80}.
%
\bibitem{Holstein81}
B.R.\ Holstein,
Phys.\ Rev.\ D {\bf 23}, 1618 (1981).
%
\bibitem{zerni} 
F.\ Zernike and H.C.\ Brinkman,
Proc.\ Kon.\ Ned.\ Acad.\ Wensch.\ {\bf 33}, 3 (1935).
%
\bibitem{F83} 
M.\ Fabre de la Ripelle,
Ann.\ Phys.\ (N.Y.) {\bf 147}, 281 (1983).
%
\bibitem{abra} 
M.\ Abramowitz and I.\ Stegun, {\it Handbook of
Mathematical Functions} (Dover Publications, Inc., New York, 1970).
%
\bibitem{Viviani05} M. Viviani, A. Kievsky, and S. Rosati,
Phys. Rev. {\bf C71}, 024006 (2005)
%
\bibitem{Fisher06} B. M. Fisher {\it et al.}, 
Phys. Rev. {\bf C74}, 034001 (2006)
%
\bibitem{HH08} 
H.M.\ Hofmann and G.M.\ Hale, Phys.\ Rev.\ C {\bf 68}, 021002(R)
(2003); Phys.\ Rev.\ C {\bf 77}, 044002 (2008).
%
\bibitem{Lazauskas09}
R.\ Lazauskas, Y-H.\ Song, and T-S.\ Park,
arXiv:0905.3119.
%
\bibitem{Deltuvapv}
A.\ Deltuva, private communication.
%
\bibitem{Zimmer02} 
O.\ Zimmer {\it et al.}, EPJdirect {\bf A1}, 1 (2002).
%
\bibitem{Huffman04} 
P.R.\ Huffman {\it et al.}, Phys.\ Rev.\ C {\bf 70}, 014004  (2004).
%
\bibitem{Ketter06} 
W.\ Ketter {\it et al.}, Eur.\ Phys.\ J.\ A {\bf 27}, 243  (2006).
%
\bibitem{Huber09}
M.G.\ Huber {\it et al.}, Phys.\ Rev.\ Lett.\ {\bf 102}, 200401 (2009);
Phys.\ Rev.\ Lett.\ {\bf 103}, 179903 (2009).
%
\bibitem{Schiavilla89}
R.\ Schiavilla, V.R.\ Pandharipande, and D.O.\ Riska,
Phys.\ Rev.\ C {\bf 40}, 2294 (1989).
%
\bibitem{Machleidt01}
R.\ Machleidt,
Phys.\ Rev.\ C {\bf  63}, 024001 (2001).
%
%
\end{thebibliography}
\end{document}